\documentclass[11pt,a4paper]{article}

\usepackage{amssymb}
\usepackage{bm}
\usepackage[dvips]{graphicx}

\begin{document}

\title{\bf The three-loop Adler $D$-function for ${\cal N}=1$ SQCD regularized by dimensional reduction}

\author{
S.S.Aleshin,\\
{\small{\em Institute for Information Transmission Problems RAS}},\\
{\small{\em 127051, Moscow, Russia}},\\
\\
A.L.Kataev,\\
{\small{\em Institute for Nuclear Research of the Russian Academy of Science,}}\\
{\small{\em 117312, Moscow, Russia}};\\
{\small{\em Moscow Institute of Physics and Technology,}}\\
{\small{\em 141700, Dolgoprudny, Moscow Region, Russia}},\\
\\
K.V.Stepanyantz\\
{\small{\em Moscow State University,}}\\
{\small{\em Faculty of Physics, Department of Theoretical Physics,}}\\
{\small{\em 119991, Moscow, Russia}}}

\maketitle

\begin{abstract}
The three-loop Adler $D$-function for ${\cal N}=1$ SQCD in the $\overline{\mbox{DR}}$ scheme is calculated starting from the three-loop result recently obtained with the higher covariant derivative regularization. For this purpose, for the theory regularized by higher derivatives we find a subtraction scheme in which the Green functions coincide with the ones obtained with the dimensional reduction and the modified minimal subtraction prescription for the renormalization of the SQCD coupling constant and of the matter superfields. Also we calculate the $D$-function in the $\overline{\mbox{DR}}$ scheme for all renormalization constants (including the one for the electromagnetic coupling constant which appears due to the SQCD corrections). It is shown that the results do not satisfy the NSVZ-like equation relating the $D$-function to the anomalous dimension of the matter superfields. However, the NSVZ-like scheme can be constructed with the help of a properly tuned finite renormalization. It is also demonstrated that the three-loop $D$-function defined in terms of the bare couplings with the dimensional reduction does not satisfy the NSVZ-like equation for an arbitrary renormalization prescription. We also investigate a possibility to present the results in the form of the $\beta$-expansion and the scheme dependence of this expansion.
\end{abstract}

\unitlength=1cm

\vspace*{-20.0cm}

\begin{flushright}
INR-TH-2019-001
\end{flushright}

\vspace*{18.5cm}


\section{Introduction}
\hspace{\parindent}

An important role in investigating strong interaction contributions to various physical quantities is played by the so-called $R$-ratio

\begin{equation}
R(s) \equiv \frac{\sigma(e^+ e^-\to \mbox{hadrons})}{\sigma_0(e^+ e^- \to \mu^+ \mu^-)} = \frac{3s}{4\pi\alpha_*^2}\, \sigma(e^+ e^-\to \mbox{hadrons}),
\end{equation}

\noindent
where $\alpha_*$ is the fine structure constant in electrodynamics and $s$ is the square of the center а mass total energy. For example, one way to obtain a value of the strong coupling constant is to compare a theoretical prediction for $R(s)$ with the experimental data \cite{Anashin:2018vdo}. Experimental and theoretical results for the $R$-ratio can be used for determining the strong interaction contribution to the muon anomalous magnetic moment coming from the hadronic vacuum polarization effects \cite{Jegerlehner:2017lbd}.

The theoretical prediction for the  $R$-ratio is closely related to the Adler $D$-function \cite{Adler:1974gd},

\begin{equation}
\bm{D}\left(\bar\alpha_s(P^2)\right) =  P^2 \int\limits_0^\infty ds\, \frac{R(s)}{(s+P^2)^2}.
\end{equation}

\noindent
Here $P$ is the Euclidean momentum, and the function $\bar\alpha_s(P^2)$ is defined as a solution of the equation $d\bar\alpha_s/d\ln P = \beta(\bar\alpha_s)$ with the boundary condition $\bar\alpha(\mu^2) = \alpha_s$, where $\mu$ denotes a scale of the renormalization (or the renormalization point).

The $D$-function allows comparing the theoretical QCD predictions with the experimental data for $R(s)$ \cite{Eidelman:1998vc}. Its theoretical expression can be defined by different ways. For example, in the region where the perturbation theory is applicable it can be found by calculating the QCD corrections to the electromagnetic coupling constant encoded in the expression

\begin{equation}\label{D_Function_Original}
\bm{D}(\alpha_s) =  - \frac{3\pi}{2}\frac{\partial}{\partial\ln P} d^{-1}(\alpha_0,\alpha_{s0},P/\Lambda)
\Big|_{\alpha\to 0} = - \frac{3\pi}{2}\frac{\partial}{\partial\ln P} d^{-1}(\alpha,\alpha_{s},P/\mu)
\Big|_{\alpha\to 0},
\end{equation}

\noindent
where $\alpha$ and $\alpha_s$ are the renormalized electromagnetic and strong coupling constants, respectively, while $\alpha_0$ and $\alpha_{s0}$ are the corresponding bare couplings. The inverse invariant charge $d^{-1}$ is related to the polarization operator $\Pi$ by the equation

\begin{equation}
d^{-1}(\alpha,\alpha_s,P/\mu) = \alpha^{-1} + 4\pi \Pi(\alpha,\alpha_s,P/\mu).
\end{equation}

\noindent
It is important that in Eq. (\ref{D_Function_Original}) the electromagnetic coupling constant is set to 0. Therefore, only quantum corrections coming from the quark and gluon loops are taken into account, while the electromagnetic field is treated as an external one,

\begin{equation}
\bm{D}(\alpha_s) = - 6\pi^2 \frac{\partial}{\partial\ln P} \Pi(\alpha \to 0,\alpha_{s},P/\mu) \equiv - 6\pi^2 \frac{\partial}{\partial\ln P} \Pi(\alpha_{s},P/\mu) ,
\end{equation}

\noindent
so that only QCD corrections survive. Graphically, this implies that the expression (\ref{D_Function_Original}) is contributed to by the diagrams with two external lines corresponding to the electromagnetic field in which all internal lines correspond to quarks and gluons (and there are no internal lines of the eletromagnetic field).

Note, that starting from the $O(\alpha_s^2)$ level the truncated perturbatve expression for the $D$-function depends on a subtraction scheme. In this paper we reveal how various renormalization prescriptions effect the $D$-function for the ${\cal N}=1$ sypersymmetric quantum chromodynamics (SQCD) interacting with the electromagnetic superfield. In this case the theoretical expression for the Adler function (\ref{D_Function_Original}) can be obtained from the massless SQCD corrections to the renormalization of the  electromagnetic coupling constant.

Another definitions of the $D$-function,

\begin{equation}\label{D_Function_Renormalized}
\widetilde D(\alpha_s) = -\frac{3\pi}{2}\frac{d}{d\ln\mu}\alpha^{-1}(\alpha_0,\alpha_{s0},\Lambda/\mu)
\Big|_{\alpha_0,\alpha_{s0} = \mbox{\scriptsize const};\ \alpha_0\to 0},
\end{equation}

\noindent
was introduced in Ref. \cite{Kataev:2017qvk}. In Appendix \ref{Appendix_D_Relation} we demonstrate that

\begin{equation}
\bm{D}(\alpha_s) = \widetilde D(\alpha_s)
\end{equation}

\noindent
in the special renormalization scheme when the SQCD-renormalization of the electromagnetic coupling constant is made with the help of the momentum (MOM)-like prescription

\begin{equation}\label{MOM_Scheme}
\Big(d^{-1}(\alpha,\alpha_s,P/\mu=1) - \alpha^{-1}\Big)\Big|_{\alpha\to 0} = 4\pi \Pi(\alpha_s,P/\mu=1) = 0.
\end{equation}

\noindent
To some extent, this equation resembles the condition $\alpha_0(\alpha,\Lambda/\mu=1) = \alpha$ used in Ref. \cite{Kataev:2013eta} for constructing the NSVZ scheme for ${\cal N}=1$ SQED. However, the boundary condition (\ref{MOM_Scheme}) is imposed to the Green function at a certain value of the momentum, while the above mentioned condition of Ref. \cite{Kataev:2013eta} is actually imposed to the renormalization constant at a certain value of the normalization point.

It is important that the prescription (\ref{MOM_Scheme}) fixes the subtraction scheme for the SQCD-renormalization of the electromagnetic coupling, while the renormalization prescription for the SQCD coupling remains unfixed.\footnote{The scheme (\ref{MOM_Scheme}) exists, because if $d^{-1}(\alpha,\alpha_s,P/\mu=1) = \alpha^{-1} + f(\alpha_s) + O(\alpha)$, where $f(\alpha_s)$ is a finite function, then Eq. (\ref{MOM_Scheme}) is satisfied for $d^{-1}(\alpha',\alpha_s,P/\mu) \equiv d^{-1}(\alpha(\alpha'),\alpha_s,P/\mu)$ with $(\alpha')^{-1} = \alpha^{-1} + f(\alpha_s)$.}

The Adler function $\bm{D}(\alpha_s)$ is known for various theories in sufficiently large orders of the perturbation theory. In QCD it has been calculated in the three-loop approximation in \cite{Chetyrkin:1979bj,Celmaster:1979xr} analytically and in \cite{Dine:1979qh} numerically; in the four-loop approximation in Ref. \cite{Gorishnii:1990vf} (see also \cite{Surguladze:1990tg,Chetyrkin:1996ez}); in the five-loop approximation in \cite{Baikov:2008jh,Baikov:2012zn}. Also the Adler function was calculated for some other models. For instance, the three-loop result for QCD supplemented by coloured scalars has been obtained in \cite{Chetyrkin:1983qc}. For ${\cal N}=1$ SQCD the two-loop Adler $D$-function has been evaluated in \cite{Kataev:1983at,Altarelli:1983pr}.

In Refs. \cite{Shifman:2014cya,Shifman:2015doa} the all-loop expression for the $D$-function of ${\cal N}=1$ SQCD has been written down. This exact equation relates the $D$-function to the anomalous dimension of the matter superfields,\footnote{Here we present this equation for the case of a general simple group $G$ and a general representation $R$, see Ref. \cite{Kataev:2017qvk}.}

\begin{eqnarray}\label{NSVZ_Like_Relation}
\widetilde D(\alpha_s) = \frac{3}{2} \sum_{\alpha=1}^{N_f}q_{\alpha}^2\, \Big(\mbox{dim}(R) - \mbox{tr}\, \widetilde\gamma(\alpha_{s})\Big),
\end{eqnarray}

\noindent
and can be considered as an analog of the exact NSVZ $\beta$-function \cite{Novikov:1983uc,Jones:1983ip,Novikov:1985rd,Shifman:1986zi} for ${\cal N}~=~1$ supersymmetric gauge theories. However, it is known that the NSVZ equation is scheme-dependent. In particular, for ${\cal N}=1$ SQED it is valid only for a certain class of the renormalization prescriptions \cite{Goriachuk:2018cac}. The general equation describing, how it changes under finite renormalizations, can be found in \cite{Kutasov:2004xu,Kataev:2014gxa}. Numerous calculations \cite{Jack:1996vg,Jack:1996cn,Jack:1998uj,Harlander:2006xq,Harlander:2009mn,Mihaila:2013wma} demonstrated that the NSVZ relation is not valid in the $\overline{\mbox{DR}}$ scheme  (i.e. for a theory regularized by the dimensional reduction \cite{Siegel:1979wq} supplemented by the modified minimal subtractions \cite{Bardeen:1978yd}). However,
investigation of supersymmetric theories revealed essential advantages of using the higher covariant derivative (HD) regularization \cite{Slavnov:1971aw,Slavnov:1972sq}  in the supersymmetric version \cite{Krivoshchekov:1978xg,West:1985jx}. In this case ${\cal N}=1$ supersymmetry is a manifest symmetry and remains unbroken in all loops. Moreover, the HD regularization allows deriving Eq. (\ref{NSVZ_Like_Relation}) in a natural and beautiful way, because for this regularization an equation similar to (\ref{NSVZ_Like_Relation}) takes place for the functions defined in terms of the bare coupling \cite{Shifman:2014cya,Shifman:2015doa},

\begin{eqnarray}\label{NSVZ_Like_Relation_Bare}
D(\alpha_{s0}) = \frac{3}{2} \sum_{\alpha=1}^{N_f}q_{\alpha}^2\, \Big(\mbox{dim}(R) - \mbox{tr}\, \gamma(\alpha_{s0})\Big),
\end{eqnarray}

\noindent
where

\begin{eqnarray}\label{D_Function_Bare}
&& D(\alpha_{s0}) = -\frac{3\pi}{2}\frac{d}{d\ln\Lambda}\alpha_0^{-1}(\alpha,\alpha_s,\Lambda/\mu)\Big|_{\alpha,\alpha_s=\mbox{\scriptsize const};\,\alpha_0\to 0};\quad\\
&& \gamma(\alpha_{s0})_i{}^j = - \frac{d}{d\ln\Lambda}\ln Z(\alpha_s,\Lambda/\mu)_i{}^j\Big|_{\alpha_s=\mbox{\scriptsize const}}.
\end{eqnarray}

\noindent
Here $\Lambda$ is the dimensionful parameter of the regularization and the condition $\alpha_0\to 0$ extracts only SQCD corrections to the electromagnetic coupling constant. In Ref. \cite{Kataev:2017qvk} Eq. (\ref{NSVZ_Like_Relation_Bare}) has been verified in the order $O(\alpha_{s0}^2)$ inclusive. By other words, the relation between the three-loop $D$-function and the two-loop anomalous dimension has been checked by explicit calculations.

The relation (\ref{NSVZ_Like_Relation_Bare}) follows from the factorization of the loop integrals giving the left hand side into integrals of double total derivatives with respect to the loop momentum. It seems to be an inherent feature of supersymmetric theories and theories with softly broken supersymmetry. The factorization into total and double total derivatives was first noted in \cite{Soloshenko:2003nc} and \cite{Smilga:2004zr} for ${\cal N}=1$ supersymmetric electrodynamics (SQED), respectively. Subsequently, a similar structure of loop integrals giving the left hand side of the NSVZ and NSVZ-like equations has been confirmed by numerous calculations for various theories \cite{Pimenov:2009hv,Stepanyantz:2011cpt,Stepanyantz:2011bz,Kazantsev:2014yna,Buchbinder:2014wra,Buchbinder:2015eva,Aleshin:2016yvj,Shakhmanov:2017soc,Kazantsev:2018nbl}. For some of them the all-loop proves have been done \cite{Stepanyantz:2011jy,Stepanyantz:2014ima,Nartsev:2016nym}.

The NSVZ and NSVZ-like equations for the renormalization group functions (RGFs) defined in terms of the bare couplings obtained with the higher derivative regularization allow constructing the all-loop renormalization prescription under which these relations are valid for RGFs defined in terms of the renormalized couplings \cite{Kataev:2013eta,Kataev:2014gxa,Kataev:2013csa,Nartsev:2016mvn,Stepanyantz:2016gtk}. Namely, it is necessary to regularize a theory by higher covariant derivatives and include into the renormalization constants only powers of $\ln\Lambda/\mu$. This prescription is analogous to the minimal subtractions \cite{Bardeen:1978yd}, so that we call it ``HD+MSL'' \cite{Kataev:2017qvk,Kazantsev:2017fdc,Stepanyantz:2017sqg}, where HD and MSL stand for Higher Derivatives and Minimal Subtraction of Logarithms, respectively. Validity of Eq. (\ref{NSVZ_Like_Relation}) in the HD+MSL scheme has been confirmed by the explicit calculation of Ref. \cite{Kataev:2017qvk} in the order $O(\alpha_s^2)$, where the scheme dependence manifests itself.

However, the Adler function (\ref{D_Function_Original}) is not certainly obtained in the HD+MSL scheme. Moreover, most calculations of the phenomenological interest in SQCD have been done in the $\overline{\mbox{DR}}$-scheme, so that it is desirable to find the result for the $D$-function in this case as well. In particular, to calculate the $D$-function (\ref{D_Function_Original}), we should use the condition (\ref{MOM_Scheme}) for the SQCD-renormalized electromagnetic coupling constant and the $\overline{\mbox{DR}}$ prescription for the renormalization of the SQCD coupling constant and the matter superfields. Fortunately, in Ref. \cite{Kataev:2017qvk} the function (\ref{D_Function_Renormalized}) has been calculated for an {\it arbitrary} renormalization prescription supplementing the higher covariant derivative regularization. The function $\bm{D}(\alpha_s)$ can be obtained from this result by a proper fixing of the renormalization scheme ambiguity. This will be done in this paper. Also we will find the result for the function (\ref{D_Function_Renormalized}) in the $\overline{\mbox{DR}}$ scheme (for SQCD coupling constant, SQCD-renormalized electromagnetic coupling constant, and the renormalization of the matter superfields). Moreover, we will investigate the scheme dependence of the $D$-function and subtraction schemes in which the NSVZ-like relation for it takes place. The results for the $D$-function will be presented in the form of the $\beta$-expansion, earlier used in the QCD case \cite{Mikhailov:2004iq,Kataev:2010du}.

The paper is organized as follows: in Sect. \ref{Section_Adler_Function_HD} we describe the theory under consideration and recall the result of Ref. \cite{Kataev:2017qvk} obtained with the higher covariant derivative regularization for an arbitrary renormalization prescription. In the next Sect. \ref{Section_Original_Adler_Function} we fix the renormalization scheme by the prescription (\ref{MOM_Scheme}) for the SQCD-renormalized electomagnetic couping constant $\alpha$ and by the $\overline{\mbox{DR}}$ prescription for the SQCD coupling constant $\alpha_s$ and the matter superfields. As a result we obtain the function $\bm{D}(\alpha_s)$ and demonstrate that it does not satisfy the NSVZ-like equation (\ref{NSVZ_Like_Relation}). Sect. \ref{Section_Adler_Function_DR} is devoted to the calculation of the $D$-function (\ref{D_Function_Renormalized}) in the case when the $\overline{\mbox{DR}}$ prescription is used for all renormalization constants. We demonstrate that the result also does not satisfy Eq. (\ref{NSVZ_Like_Relation}). Moreover, in this section we calculate the function $D(\alpha_0)$ (defined in terms of the bare couplings) for the theory regularized by dimensional reduction and show that (unlike the case of using the higher covariant derivative regularization) it does not satisfy the NSVZ-like relation (\ref{NSVZ_Like_Relation_Bare}). The scheme dependence of the $D$-function is investigated in Sect. \ref{Section_Scheme_Dependence}, where we obtain the equation describing how the $D$-function changes under the finite renormalizations. Using this equation in Sect. \ref{Section_NSVZ} we discuss various ways to restore Eq. (\ref{NSVZ_Like_Relation_Bare}) for the three-loop $D$-function in the case of using the regularization by dimensional reduction. In particular, we demonstrate that even the function $\bm{D}(\alpha_s)$ can satisfy this relation for a proper choice of a prescription for the renormalization of the matter superfields. In Sect. \ref{Section_Beta_Expansion} the results for the $D$-function are presented in the form of the $\beta$-expansion. In this section we also discuss the dependence of the $\beta$-expansion on a renormalization prescription in the considered approximation and collect the scheme dependent coefficients for various subtraction schemes in Table \ref{Table_Scheme_Dependence}.

\section{The three-loop expression for $\widetilde D(\alpha_s)$ for ${\cal N}=1$ SQCD regularized by higher derivatives}
\hspace{\parindent}\label{Section_Adler_Function_HD}

In this paper we consider massless ${\cal N}=1$ SQCD (with a simple gauge group $G$ and the matter superfields in a certain representation) interacting with the external electromagnetic field in the supersymmetric way. Such a theory is invariant under the $G\times U(1)$ gauge transformations, and, therefore, contains two coupling constants $g$ and $e$ corresponding to the factors $G$ and $U(1)$, respectively. It is convenient to write the action of this theory in terms of the ${\cal N}=1$ superfields, because in this case ${\cal N}=1$ supersymmetry is manifest,

\begin{eqnarray}\label{Action_SQCD_Classical}
&& S = \frac{1}{2 g_0^2}\,\mbox{Re}\,\mbox{tr}\int d^4x\, d^2\theta\,W^a W_a
+ \frac{1}{4 e_0^2}\,\mbox{Re}\int d^4x\, d^2\theta\,\bm{W}^a \bm{W}_a \nonumber\\
&&\qquad\qquad\quad
+\sum_{\alpha=1}^{N_{f}} \frac{1}{4}\,\int d^4x\, d^4\theta\,\Big(\phi_{\alpha}^{+}e^{2V+2q_{\alpha}\bm{V}}\phi_{\alpha}
+ \widetilde\phi_{\alpha}^{+}e^{-2V^t-2q_{\alpha}\bm{V}}\widetilde\phi_{\alpha}\Big).
\end{eqnarray}

\noindent
Here $g_0$ and $e_0$ are the bare coupling constants. (Also we will use the notations $\alpha_{s0}\equiv g_0^2/4\pi$ and $\alpha_0 \equiv e_0^2/4\pi$.) The non-Abelian gauge superfield corresponding to the subgroup $G$ is denoted by $V$, while $\bm{V}$ is the external Abelian gauge superfield corresponding to the subgroup $U(1)$. The corresponding gauge superfield strengths are defined as

\begin{eqnarray}
W_a=\frac{1}{8}\bar D^2(e^{-2V} D_a e^{2V}); \qquad \bm{W}_a = \frac{1}{4}\bar D^2 D_a \bm{V}.
\end{eqnarray}

\noindent
The chiral matter superfields $\phi_{\alpha}$, $\widetilde{\phi}_{\alpha}$ in the representations $R$ and $\bar{R}$ with the electric charges $q_\alpha$ and $-q_\alpha$, respectively, describe Dirac fermions and their scalar superpartners.

The three-loop $D$-function (\ref{D_Function_Renormalized}) for this theory has been calculated in Ref. \cite{Kataev:2017qvk} in the case of using the supersymmetric version of the higher covariant derivative regularization. The main idea of this regularization \cite{Slavnov:1971aw,Slavnov:1972sq} is to add a higher derivative term to the action, so that the divergences remain only in the one-loop order. These remaining divergences are regularized by inserting relevant Pauli--Villars determinants into the generating functional \cite{Slavnov:1977zf}. The masses of the Pauli--Villars superfields should be proportional to the dimensionful parameter $\Lambda$ in the higher derivative term. This ensures that the regularized theory contains the only dimensionful parameter.

For the theory (\ref{Action_SQCD_Classical}) it is possible to use two sets of the Pauli--Villars superfields: three chiral superfields $\varphi_a$ in the adjoint representation of the gauge group and $N_f$ anticommuting chiral superfields $\Phi_\alpha$ and $\widetilde\Phi_\alpha$ with the same quantum numbers as $\phi_\alpha$ and $\widetilde\phi_\alpha$, respectively. The former superfields cancel one-loop divergences introduced by the gauge superfield $V$ and ghosts, while the latter ones cancel one-loop divergences coming from the matter loop. Their masses are related to the parameter $\Lambda$ in the higher derivative term by the equations

\begin{equation}
M_\varphi = a_\varphi\Lambda;\qquad M = a\Lambda,
\end{equation}

\noindent
where $a_\varphi$ and $a$ are coupling-independent parameters. Their values can of course be fixed by a certain prescription, if necessary.

The result for the three-loop $D$-function obtained in \cite{Kataev:2017qvk} with the help of the higher covariant derivative regularization (for the regulator function $R(x)= 1+ x^n$, in the Feynman gauge, see Ref. \cite{Kataev:2017qvk} for details) can be written as

\begin{eqnarray}\label{D_Function_HD}
&& \widetilde D(\alpha_s)=
\frac{3}{2}\sum_{\alpha=1}^{N_{f}}q_{\alpha}^2\,\Big\{\mbox{dim}(R)+\frac{\alpha_s}{\pi}\mbox{tr}\,C(R)+\frac{\alpha^2_s}{\pi^2}
\Big[\,\frac{3}{2}C_2\mbox{tr}\,C(R)\Big(\ln a_{\varphi}+1+d_2-b_{11}\Big)\qquad\nonumber\\
&& - N_{f}T(R)\,\mbox{tr}\, C(R)\Big(\ln a+1+d_2-b_{12}\Big)
-\frac{1}{2}\mbox{tr}\,\Big(C(R)^2\Big)\Big]\Big\} + O(\alpha_s^3).
\end{eqnarray}

\noindent
In this equation $\alpha_s = g^2/4\pi$ is the renormalized SQCD coupling constant,

\begin{eqnarray}
&& \mbox{tr}\,(T^{A}T^{B}) = T(R)  \delta^{AB};\qquad\ C(R)_i{}^j = (T^AT^A)_i{}^j;\qquad\nonumber\\
&& C_2\delta^{CD} = f^{ABC}f^{ABD};\qquad\quad r=\delta^{AA} = \mbox{dim} G,
\end{eqnarray}

\noindent
where $T^A$ and $f^{ABC}$ are the generators of the representation $R$ and the structure constants, respectively. In our notation the generators of the fundamental representation are supposed to be normalized by the condition $\mbox{tr}(t^A t^B) = \delta^{AB}/2$.

It is important that the result (\ref{D_Function_HD}) is valid for an arbitrary subtraction scheme characterized by the finite constants $b_{11}$, $b_{12}$, and $d_2$. Fixing the renormalization prescription we fix values of these constants. To be exact, the constants $b_{11}$ and $b_{12}$ appear in the one-loop approximation in the equation relating the bare ($\alpha_{s0}=g_0^2/4\pi$) and renormalized ($\alpha_s = g^2/4\pi$) SQCD coupling constants,

\begin{eqnarray}\label{AlphaS_Renormalization}
\frac{1}{\alpha_{0s}}= \frac{1}{\alpha_{s}} + \frac{1}{2\pi}\Big[3C_2\Big(\ln\frac{\Lambda}{\mu}+b_{11}\Big)-2N_fT(R)\Big(\frac{\Lambda}{\mu}+b_{12}\Big)\Big] + O(\alpha_s),
\end{eqnarray}

\noindent
where $\mu$ is a renormalization point. Similarly, the finite constants $d_1$ and $d_2$ determine the relation between the bare ($\alpha_0=e_0^2/4\pi$) and SQCD-renormalized ($\alpha = e^2/4\pi$) electromagnetic coupling constants in the two-loop order,

\begin{eqnarray}\label{Alpha_Renormalization}
\frac{1}{\alpha_{0}} = \frac{1}{\alpha} - \frac{1}{\pi}\sum_{\alpha=1}^{N_{f}}q_{\alpha}^2\, \mbox{dim}(R)
\Big(\ln\frac{\Lambda}{\mu}+d_1\Big) - \frac{\alpha_s}{\pi^2}\sum_{\alpha=1}^{N_{f}}q_{\alpha}^2\, \mbox{tr}\,C(R)\Big(\ln\frac{\Lambda}{\mu}+d_2\Big)+O(\alpha_s^2).
\end{eqnarray}

\noindent
And again we stress that $d_1$ and $d_2$ enter only in the SQCD part of the electromagnetic coupling constant renormalization.

\section{The three-loop Adler function $\bm{D}(\alpha_s)$ in the $\overline{\mbox{DR}}$ scheme for the SQCD coupling constant and the matter superfields}
\hspace{\parindent}\label{Section_Original_Adler_Function}

First, we would like to obtain the expression for the Adler function $\bm{D}(\alpha_s)$ defined by Eq. (\ref{D_Function_Original}) in the case when the theory is regularized by dimensional reduction and the $\overline{\mbox{DR}}$ renormalization prescription is used for the renormalization of the SQCD coupling constant and of the matter superfields. Certainly, this expression can be obtained from Eq. (\ref{D_Function_HD}), which is valid for an arbitrary choice of a subtraction scheme. In the case we are interested in this section it is necessary to find such values of the constants $b_{11}$, $b_{12}$, and $d_2$ that correspond to the considered renormalization prescription, i.e. $\overline{\mbox{DR}}$ scheme for the SQCD coupling constant and the matter renormalization, and MOM-like Eq. (\ref{MOM_Scheme}) for the SQCD-renormalized electromagnetic coupling constant.

Following Ref. \cite{Kataev:2017qvk}, we find the values of $b_{11}$ and $b_{12}$ (defining the renormalization of the SQCD coupling constant) by comparing the expressions for the two-loop anomalous dimension of the matter superfields calculated with the higher covariant derivatives and in the $\overline{\mbox{DR}}$ scheme. This gives the equations \cite{Kataev:2017qvk}

\begin{equation}\label{B_Relation}
b_{11} = \ln a_{\varphi}+g_1+\frac{1}{2};\qquad  b_{12}=\ln a+g_1+\frac{1}{2},
\end{equation}

\noindent
where $g_1$ is a finite constant entering the one-loop renormalization of the matter superfields,

\begin{equation}\label{Z_HD}
\ln Z_i{}^j = \frac{\alpha_s}{\pi} C(R)_i{}^j\Big(\ln\frac{\Lambda}{\mu} + g_1\Big) + O(\alpha_s^2).
\end{equation}

\begin{figure}[h]
\begin{picture}(0,2)
\put(4.2,0.1){\includegraphics[scale=0.4]{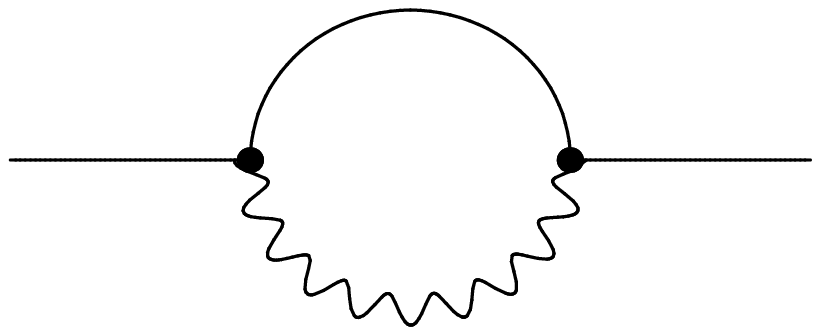}}
\put(8.2,-0.2){\includegraphics[scale=0.4]{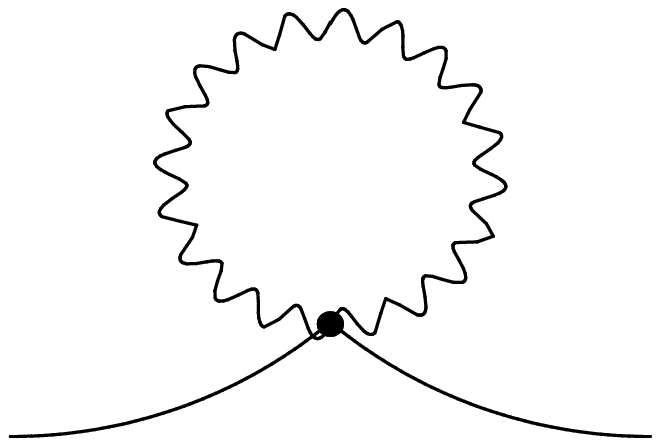}}
\end{picture}
\caption{The superdiagrams giving the two-point Green function of the matter superfields in the one-loop approximation.}\label{Figure_Matter_2Point}
\end{figure}

Therefore, we also need to know the constant $g_1$. It can be found by comparing the one-loop renormalized two-point Green functions of the matter superfields written in terms of the renormalized couplings calculated with the higher covariant derivative regularization and with dimensional reduction supplemented by the modified minimal subtractions. A part of the effective action corresponding to this Green function can be written in terms of the renormalized quantities as

\begin{equation}\label{Effective_Action_Phi}
\Gamma^{(2)}_\phi = \frac{1}{4}\sum_{\alpha=1}^{N_{f}} \int
\frac{d^4p}{(2\pi)^4}\, d^4\theta\, \Big(\phi^{*i}_{R\, \alpha}(\theta,-p)\,\phi_{R\, \alpha j}(\theta,p) + \widetilde\phi^{*}_{R\, \alpha j}(\theta,-p)\,\widetilde\phi^i_{R\, \alpha}(\theta,p)\Big)\, (ZG)_i{}^j(\alpha_{s},p/\mu),
\end{equation}

\noindent
where the subscript $R$ marks the renormalized superfields. In the one-loop approximation the contribution to the function $(G_R)_i{}^j = (ZG)_i{}^j(\alpha_s,p/\mu)$ is given by the superdiagrams presented in Fig. \ref{Figure_Matter_2Point}. In the case of using the higher derivative regularization (in the Feynman gauge with the regulator $R(x)=1+x^n$) the result can be written as

\begin{equation}\label{G_Integral_HD}
\big(G_{R,\,\mbox{\scriptsize HD}}\big)_i{}^{j} = \lim\limits_{\Lambda\to \infty}\, Z_i{}^k \Big(\delta_k^j -8\pi\alpha_{s0,\,\mbox{\scriptsize HD}}\,  C(R)_k{}^j \int\frac{d^4k}{(2\pi)^d}\frac{1}{k^2 R_k (k+p)^2} + O(\alpha_{s0,\,\mbox{\scriptsize HD}}^2)\Big),
\end{equation}

\noindent
where (for the considered regulator function) $R_k = 1+k^{2n}/\Lambda^{2n}$. Using the result for the integral in Eq. (\ref{G_Integral_HD}) calculated in Refs. \cite{Soloshenko:2002np,Soloshenko:2003sx} and substituting the renormalization constant $Z_i{}^k$ from Eq. (\ref{Z_HD}), we obtain

\begin{equation}\label{G_Renormalized_HD}
\big(G_{R,\,\mbox{\scriptsize HD}}\big)_i{}^{j} = \delta_i^j - \frac{\alpha_s}{\pi} C(R)_i{}^j \Big(\ln\frac{\mu}{P}+\frac{1}{2}-g_1\Big) + O(\alpha_s^2).
\end{equation}

From the other side, calculating the considered Green function with the help of dimensional reduction, we obtain

\begin{eqnarray}\label{G_Integral_DR}
&& \big(G_{R,\,\overline{\mbox{\scriptsize DR}}}\big)_i{}^{j} = \lim\limits_{\Lambda_{\mbox{\tiny DR}}\to \infty;\, \varepsilon\to 0}\, (Z_{\overline{\mbox{\scriptsize DR}}})_i{}^k\Big(\delta_k^j - 8\pi\alpha_{s0,\overline{\mbox{\scriptsize DR}}}\,  C(R)_k{}^j (\Lambda_{\mbox{\scriptsize DR}})^{\varepsilon} \int\frac{d^dk}{(2\pi)^d}\frac{1}{k^2 (k+p)^2}\qquad\nonumber\\
&& + O(\alpha_{s0,\overline{\mbox{\scriptsize DR}}}^2)\Big),
\end{eqnarray}

\noindent
where $\varepsilon \equiv 4-d$ and $\Lambda_{\mbox{\scriptsize DR}}$ is the dimensionful parameter of the regularized theory.\footnote{Note that in the case of using dimensional reduction the tadpole diagram in Fig. \ref{Figure_Matter_2Point} vanishes and does not contribute to Eq. (\ref{G_Integral_DR}).} Introducing the parameter $\Lambda_{\mbox{\scriptsize DR}}$ makes the coupling constant $e_0$ dimensionless. In this paper we do not follow the usual convention $\Lambda_{\mbox{\scriptsize DR}} = \mu$. This is convenient, because then the results obtained with the dimensional technique will look similar to the ones found with the higher covariant derivative regularization, see \cite{Aleshin:2015qqc,Aleshin:2016rrr}. By construction, in the version of the $\overline{\mbox{DR}}$ scheme adopted in this paper the renormalization constant $(Z_{\overline{\mbox{\scriptsize DR}}})_i{}^k$ contains only $\varepsilon$-poles and powers of $\ln\bar\Lambda/\mu$, where

\begin{equation}\label{Bar_Lambda}
\bar\Lambda \equiv \Lambda_{\mbox{\scriptsize DR}}\exp(-\gamma/2)\sqrt{4\pi}.
\end{equation}

\noindent
Using this prescription, we obtain

\begin{eqnarray}\label{Z_DR}
&& \big(Z_{\overline{\mbox{\scriptsize DR}}}\big)_i{}^j = \delta_i^j + \frac{\alpha_s}{\pi} C(R)_i{}^j\Big(\frac{1}{\varepsilon} + \ln\frac{\bar\Lambda}{\mu}\Big) + O(\alpha_s^2);\\
\label{G_Renormalized_DR}
&& \big(G_{R,\,\overline{\mbox{\scriptsize DR}}}\big)_i{}^{j} = \delta_i^j - \frac{\alpha_{s}}{\pi} C(R)_{i}^{\,j} \Big(\ln\frac{\mu}{P}+1\Big) + O(\alpha_s^2).
\end{eqnarray}

However, the renormalized all-order perturbative expansion of the Green function should not depend on the regularization and renormalization details,

\begin{equation}\label{Renormalized_G_Equality}
\big(G_{R,\,\mbox{\scriptsize HD}}\big)_i{}^{j} = \big(G_{R,\,\overline{\mbox{\scriptsize DR}}}\big)_i{}^{j}.
\end{equation}

\noindent
The Green functions entering the left and right hand sides this equation are given by Eqs. (\ref{G_Renormalized_HD}) and (\ref{G_Renormalized_DR}), respectively. From Eq. (\ref{Renormalized_G_Equality}) we obtain such a value of the constant $g_1$ for which the corresponding renormalization prescription (with the higher derivative regularization) is equivalent to the $\overline{\mbox{DR}}$ scheme,

\begin{equation}\label{G1_Result}
g_1 = -\frac{1}{2}.
\end{equation}

\noindent
This value was calculated by equating the finite parts of the considered Green functions. Substituting it to Eq. (\ref{B_Relation}), we obtain

\begin{equation}\label{B_Values}
b_{11} = \ln a_{\varphi};\qquad b_{12}=\ln a.
\end{equation}

Also Eq. (\ref{D_Function_HD}) contains the finite constant $d_2$ which determines the renormalization scheme for the SQCD-renormalized electromagnetic coupling constant. It can be found by a similar method, namely, by comparing the renormalized two-point Green functions of the Abelian gauge superfield $\bm{V}$ in the two-loop approximation. The corresponding part of the effective action has the form

\begin{equation}\label{Effective_Action_V}
\Gamma^{(2)}_{\bm{V}} = - \frac{1}{16\pi} \int
\frac{d^4p}{(2\pi)^4}\,d^4\theta\,\bm{V}(\theta,-p)\,\partial^2\Pi_{1/2}
\bm{V}(\theta,p)\, d^{-1}(\alpha,\alpha_{s},p/\mu),
\end{equation}

\noindent
where the supersymmetric transverse projection operator is denoted by $\Pi_{1/2}\equiv - D^a \bar D^2 D_a/8 = -\bar D^{\dot a} D^2 \bar D_{\dot a}/8$. Note that in calculating the Adler $D$-function the Abelian superfield $\bm{V}$ is considered as the external one, so that the function $d^{-1}$ depends on $\alpha_0$ only in the tree approximation. This implies that

\begin{eqnarray}
d^{-1}(\alpha,\alpha_s,P/\mu) = \alpha^{-1}+4\pi\Pi(\alpha_{s},P/\mu),
\end{eqnarray}

\noindent
where the polarization operator $\Pi$ is independent of $\alpha$, and $P$ is the Euclidean momentum. For the considered theory it can be equivalently defined by the equation

\begin{equation}
\langle J(x_1,\theta_1) J(x_2,\theta_2) \rangle = \frac{i}{2}\int \frac{d^4p}{(2\pi)^4}\, \Pi(\alpha_s,p/\mu)\, \partial^2\Pi_{1/2}\delta^4(\theta_1-\theta_2) \exp\left(-ip_\alpha(x_1^\alpha - x_2^\alpha)\right),
\end{equation}

\noindent
where the corresponding SQCD current superfield is

\begin{equation}
J = \frac{1}{2} \sum\limits_{\alpha=1}^{N_f} q_\alpha \Big(\phi^+_\alpha e^{2V} \phi_\alpha - \widetilde\phi^+_\alpha e^{-2V^t} \widetilde\phi_\alpha\Big).
\end{equation}

\begin{figure}[h]
\begin{picture}(0,5)
\put(0.7,2.9){\includegraphics[scale=0.16]{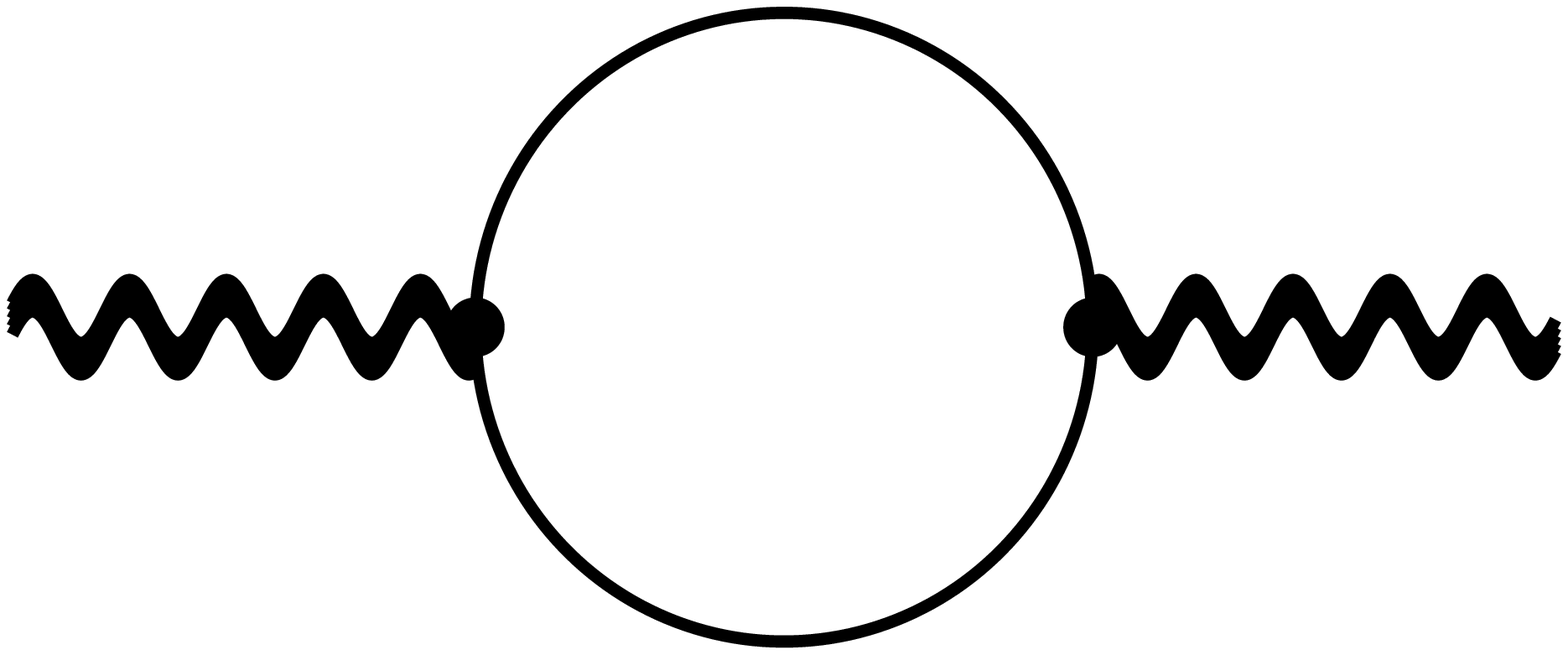}}
\put(5.05,2.6){\includegraphics[scale=0.16]{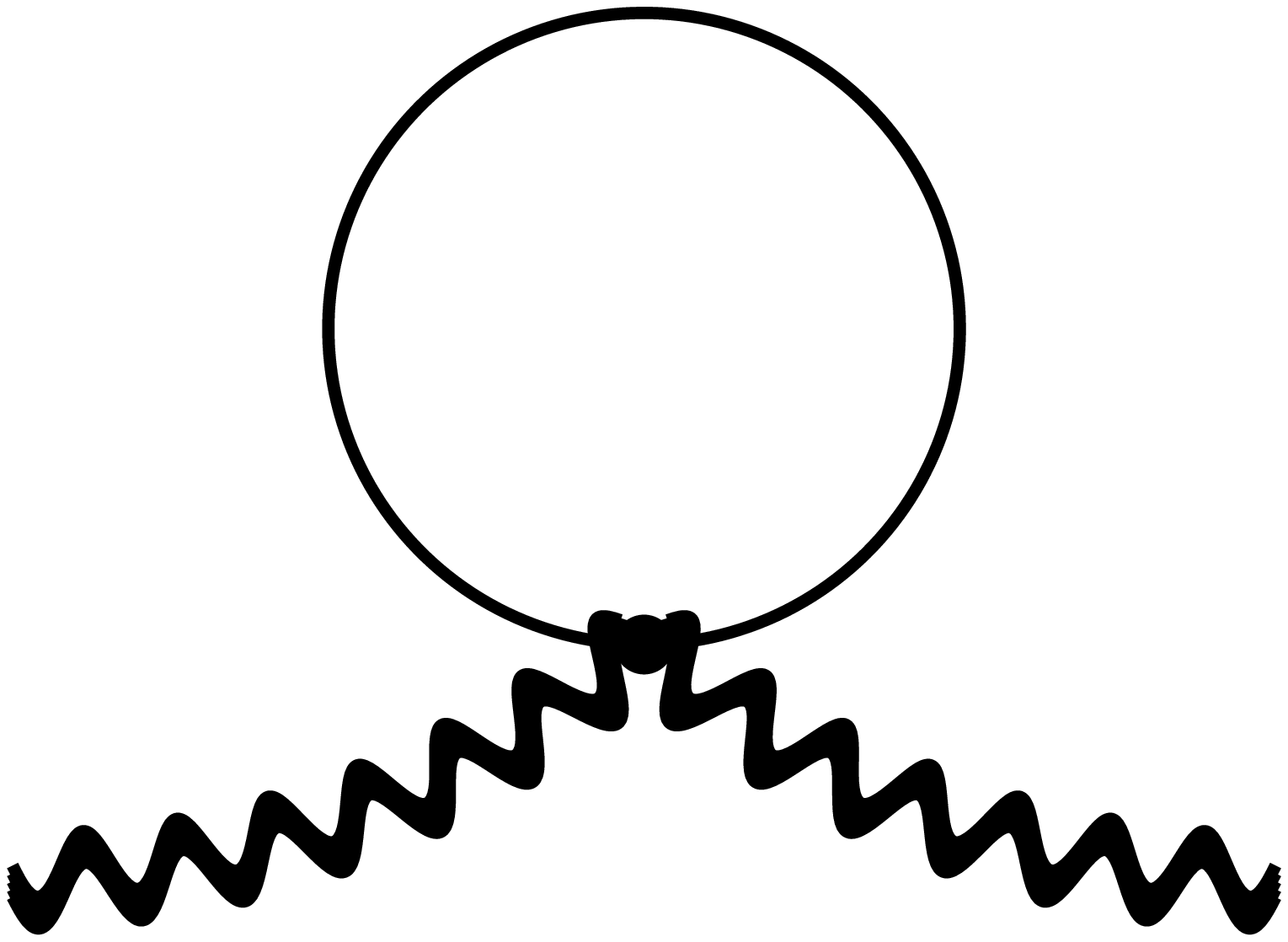}}
\put(8.7,2.8){\includegraphics[scale=0.16]{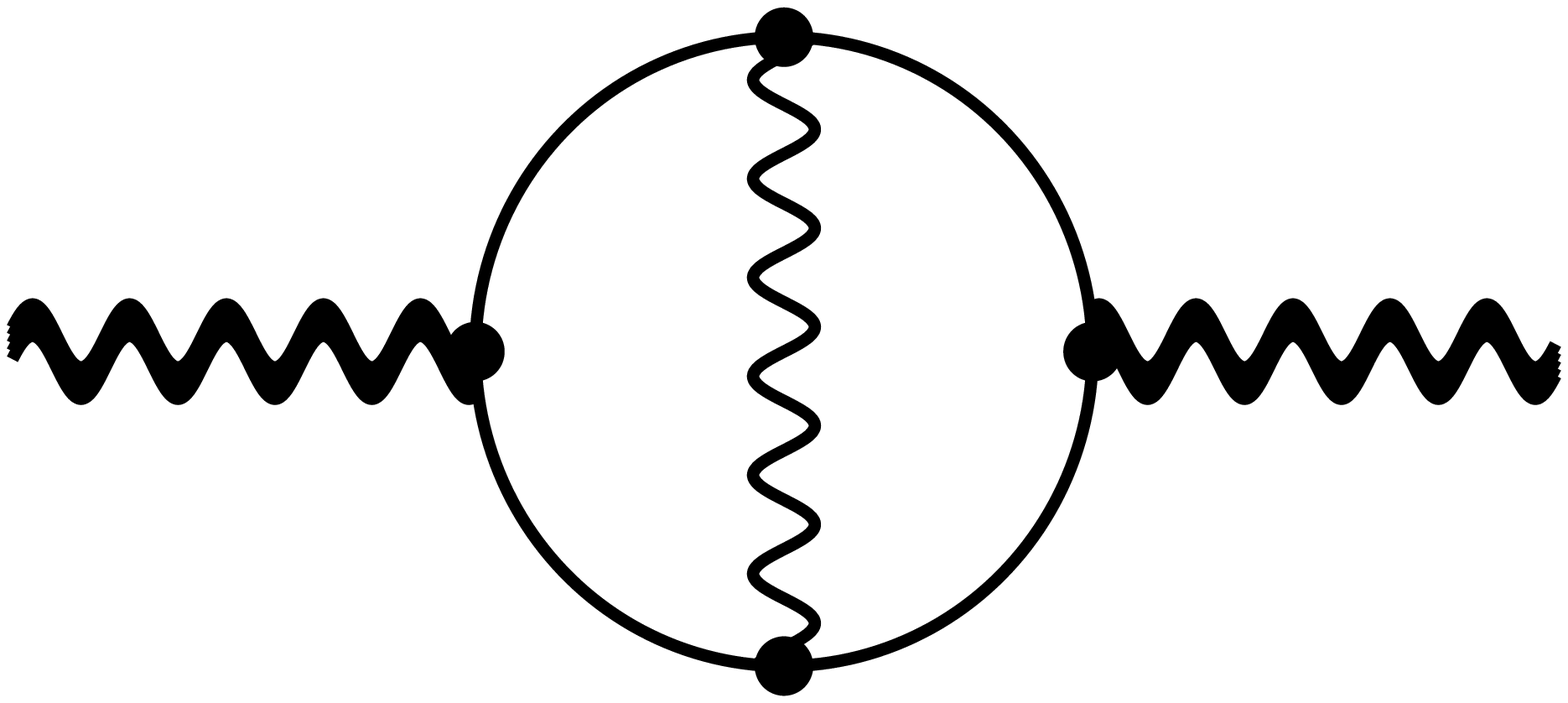}}
\put(12.7,2.8){\includegraphics[scale=0.16]{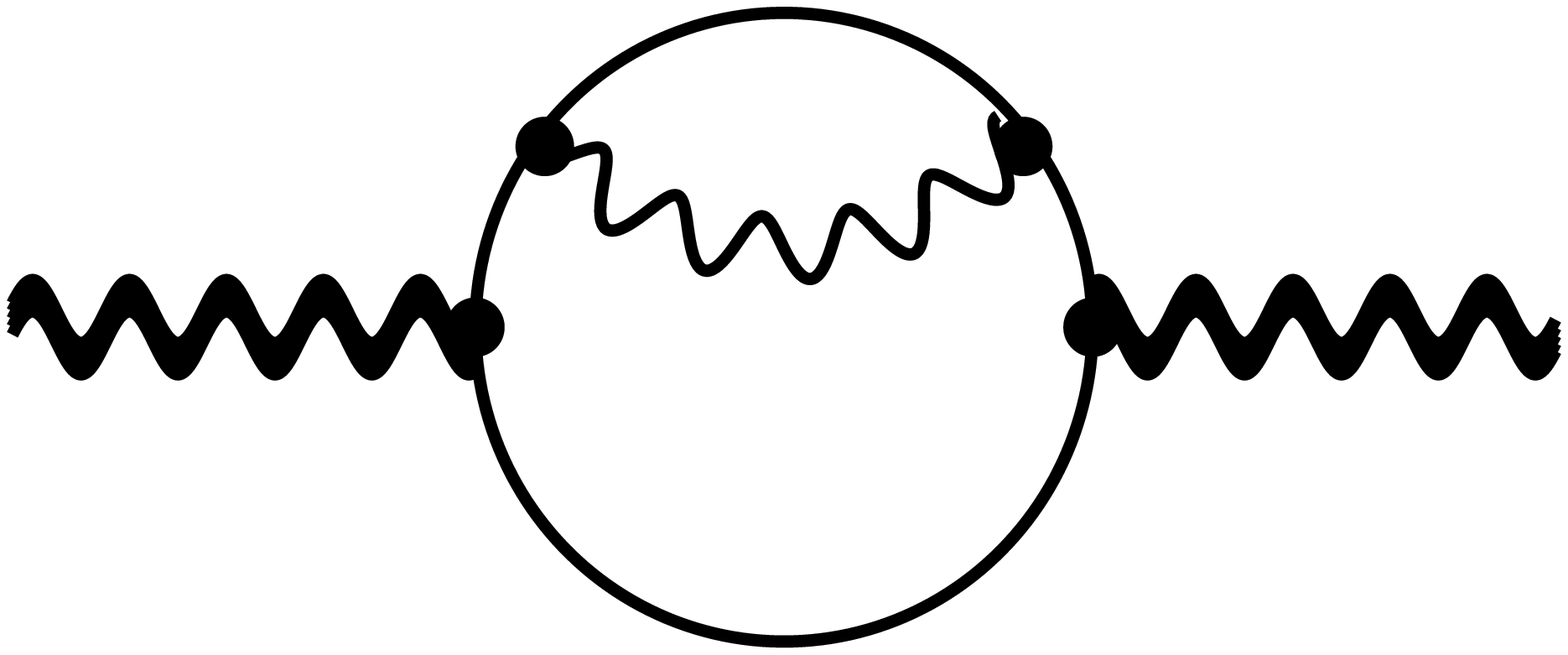}}
\put(0.7,0.2){\includegraphics[scale=0.16]{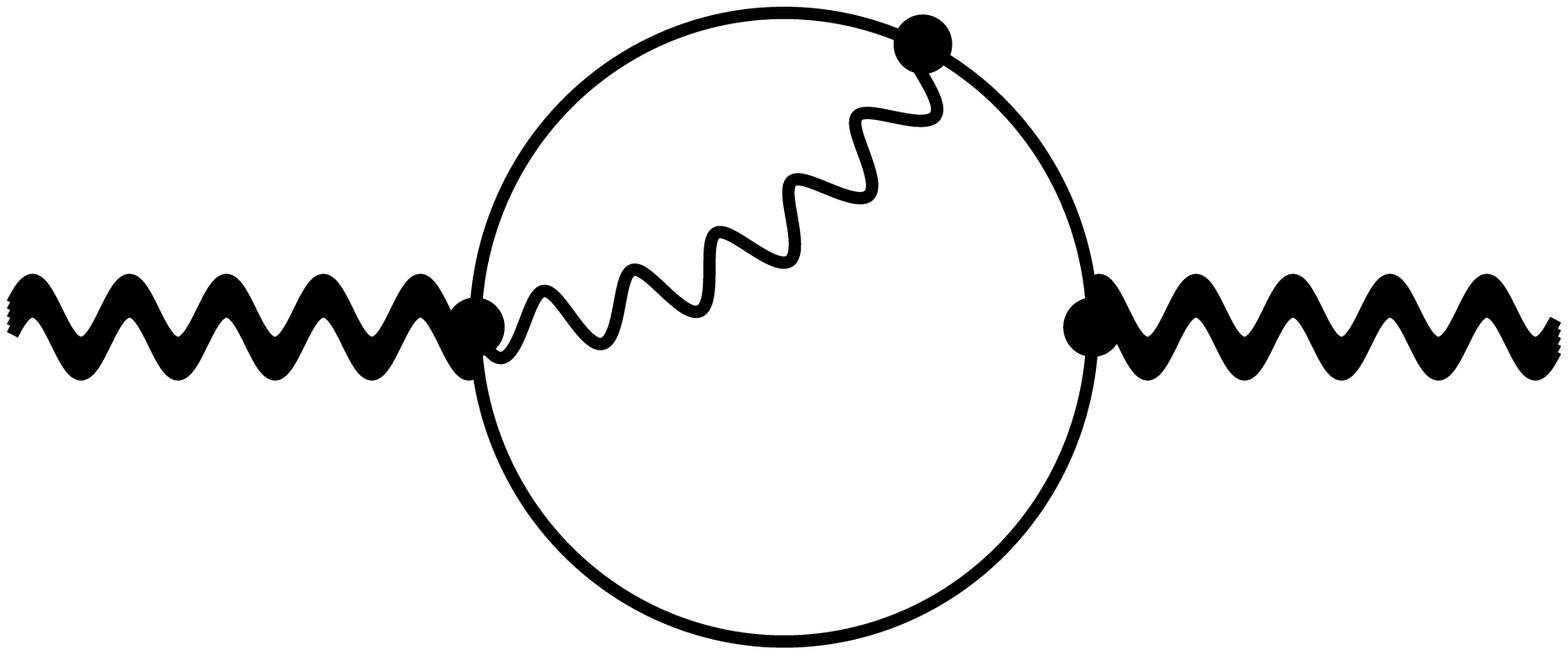}}
\put(4.8,0.2){\includegraphics[scale=0.16]{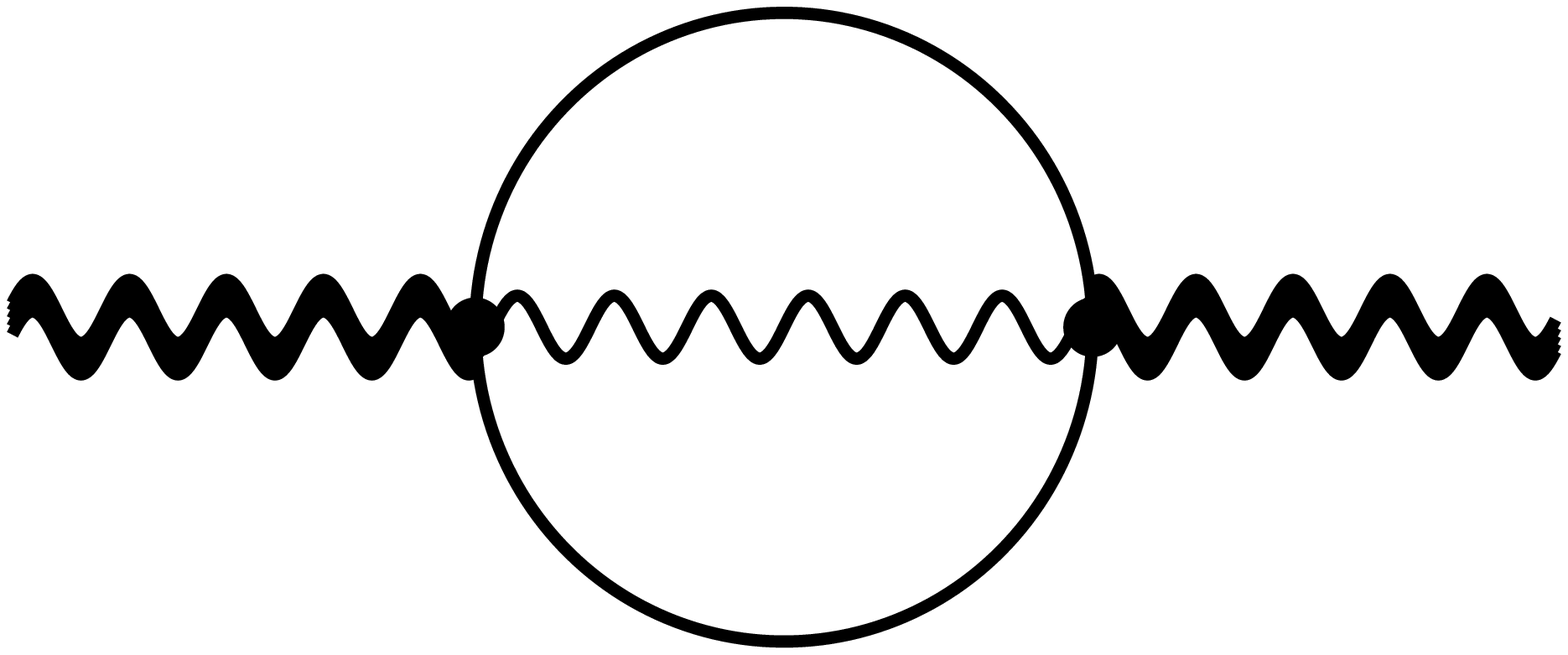}}
\put(9.1,0){\includegraphics[scale=0.16]{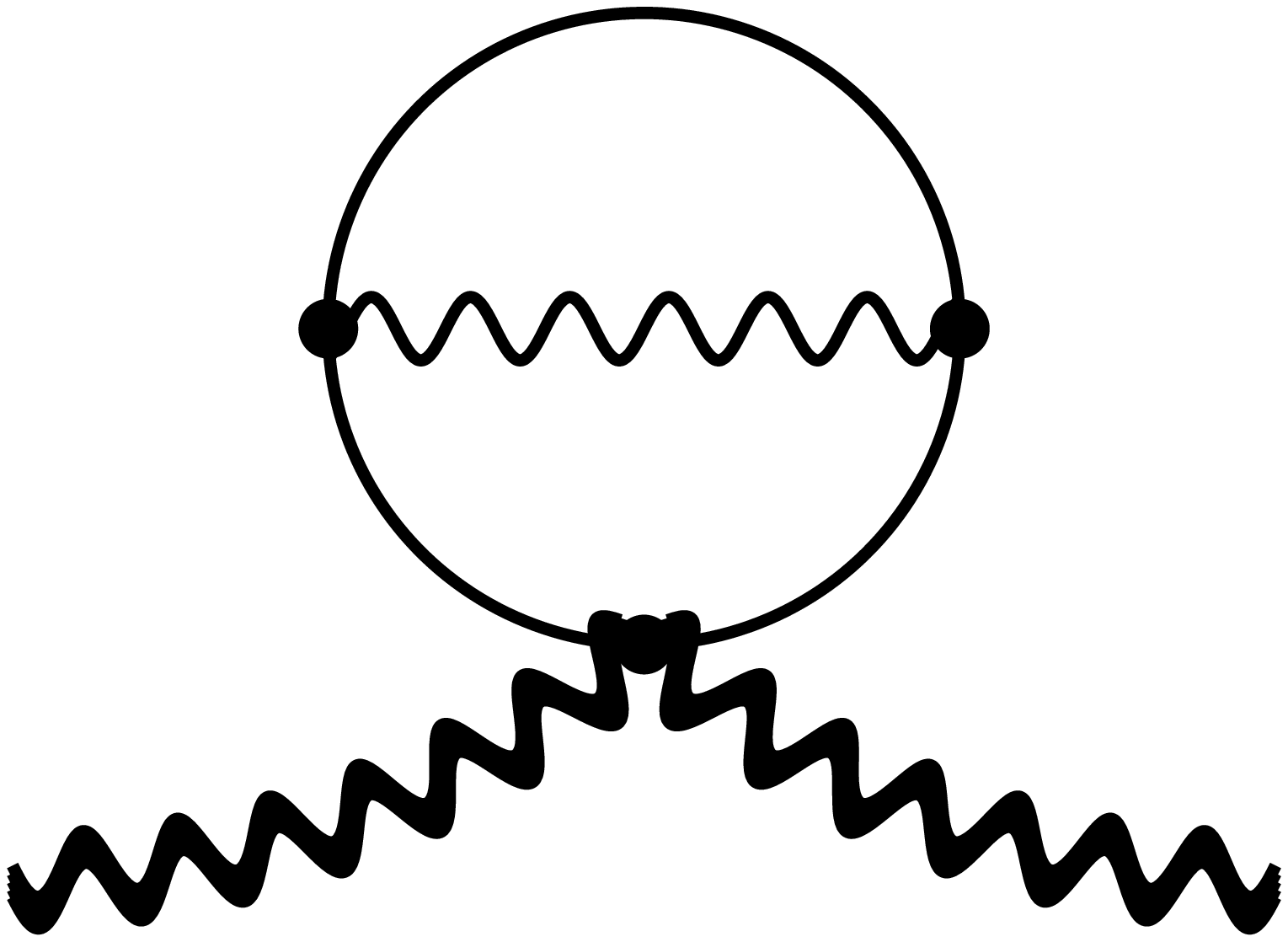}}
\put(13.1,0){\includegraphics[scale=0.16]{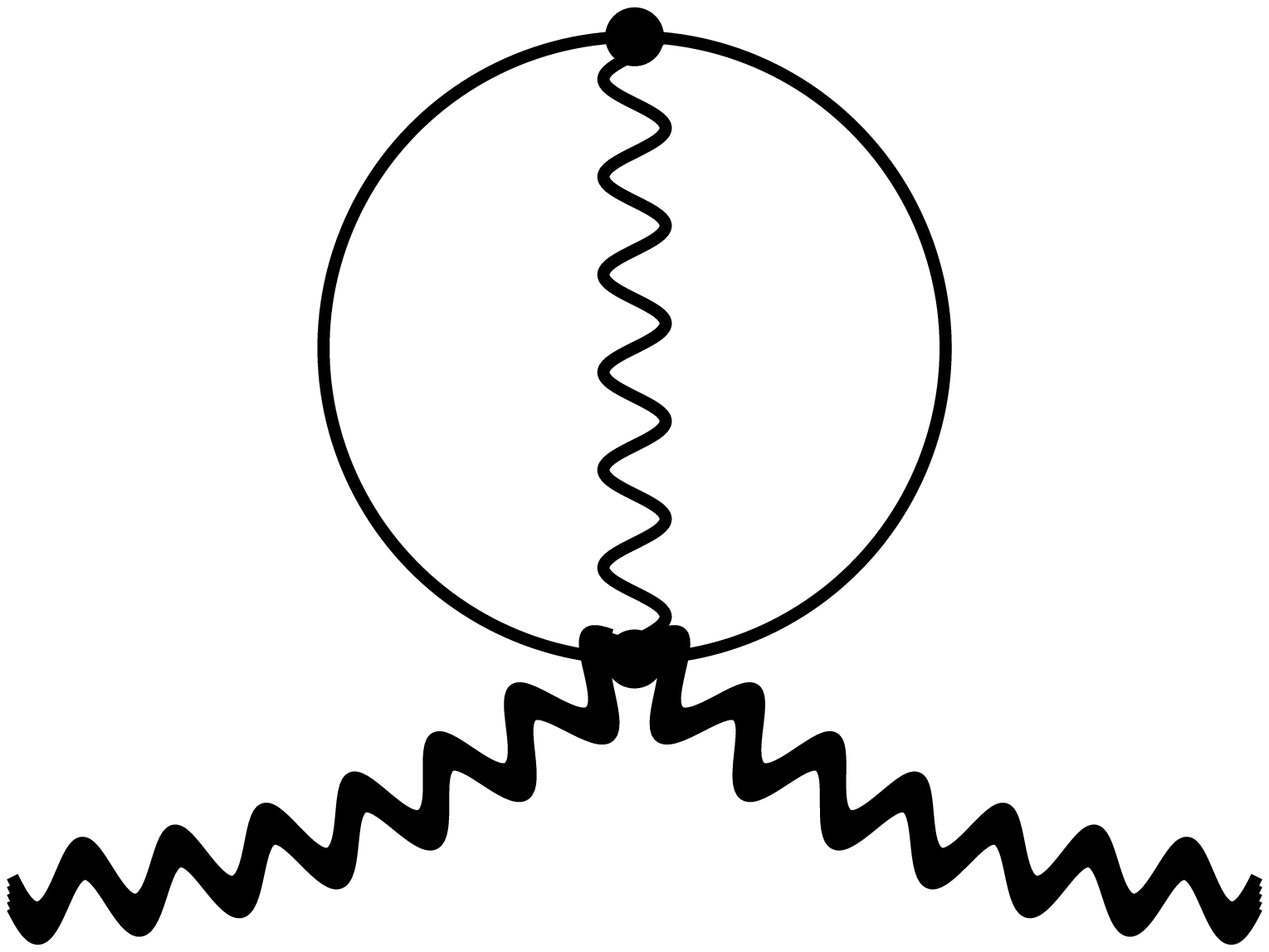}}
\put(0.8,4.2){$(1)$} \put(4.8,4.2){$(2)$}
\put(8.8,4.2){$(3)$} \put(12.8,4.2){$(4)$}
\put(0.8,1.5){$(5)$} \put(4.8,1.5){$(6)$}
\put(8.8,1.5){$(7)$} \put(12.8,1.5){$(8)$}
\end{picture}
\caption{The superdiagrams giving the two-point Green function of the Abelian gauge superfield $\bm{V}$ in the two-loop approximation.}\label{Figure_Gauge_2Point}
\end{figure}

To obtain the expressions for the function $d^{-1}$ in the considered approximation, it is necessary to calculate the superdiagrams presented in Fig. \ref{Figure_Gauge_2Point}. In this figure the bold external wavy lines correspond to the electromagnetic superfield $\bm{V}$, the solid lines denote matter propagators, and the usual wavy lines denote propagators of the non-Abelian gauge superfield $V$. Note that the result can be obtained from the corresponding result for ${\cal N}=1$ SQED with $N_f$ flavors by multiplying the latter to the factor $\sum_\alpha q_\alpha^2 \dim(R)/N_f$ in the one-loop approximation (the diagrams (1) and (2)) and to factor $\sum\limits_\alpha q_\alpha^2\,\mbox{tr}\, C(R)/N_f$ in the two-loop approximation (the diagrams (3) -- (8)).\footnote{For the version of the higher covariant derivative regularization used in Ref. \cite{Kataev:2017qvk} the interaction vertices in the diagrams presented in Fig. \ref{Figure_Gauge_2Point} do not contain higher derivative terms.} In the case of using the higher covariant derivative regularization this prescription allows obtaining the function $d^{-1}$ expressed in terms of the renormalized quantities from the ${\cal N}=1$ SQED result which can be found by combining the results presented in Refs. \cite{Kataev:2013eta,Kataev:2013csa} (derived on the base of the calculation made in Ref. \cite{Soloshenko:2003nc}),

\begin{eqnarray}\label{DInverse_Function_HD}
&& d^{-1}_{\mbox{\scriptsize HD}} = \alpha^{-1}+\frac{1}{\pi}\sum_{\alpha=1}^{N_{f}}q_{\alpha}^2\, \mbox{dim}(R)\Big(\ln\frac{\mu}{P} +\ln a +1 -d_1\Big)\nonumber\\
&&\qquad\qquad\qquad\qquad\quad +\frac{\alpha_{s}}{\pi^2}\sum_{\alpha=1}^{N_{f}}q_{\alpha}^2\, \mbox{tr}\,C(R)\Big(\ln\frac{\mu}{P}+\frac{3}{2}-\frac{3}{2}\zeta(3)-d_2\Big) + O(\alpha_s^2).\qquad
\end{eqnarray}

\noindent
In this equation the finite constants $d_1$ and $d_2$ originate from the SQCD-renormalization of the electromagnetic coupling constant, see Eq. (\ref{Alpha_Renormalization}). Their values should be fixed by the boundary condition (\ref{MOM_Scheme}). Setting $\mu=P$ and equating the coefficients of $\alpha_s$ powers, we find

\begin{equation}\label{D_Values_MOM}
d_1 = \ln a +1;\qquad d_2 = \frac{3}{2} - \frac{3}{2}\zeta(3).
\end{equation}

\noindent
Substituting the finite constants (\ref{B_Values}) and (\ref{D_Values_MOM}) into Eq. (\ref{D_Function_HD}) we obtain the function $\bm{D}(\alpha_s)$ in the case when the renormalization of the SQCD coupling constant and of the matter superfields is made by the $\overline{\mbox{DR}}$ prescription,

\begin{eqnarray}\label{D_Original_Three_Loop}
&& \bm{D}(\alpha_s)=
\frac{3}{2}\sum_{\alpha=1}^{N_{f}}q_{\alpha}^2\,\Big\{\mbox{dim}(R)+\frac{\alpha_s}{\pi}\mbox{tr}\,C(R)+\frac{\alpha^2_s}{\pi^2}
\Big[\, - \frac{1}{2}\mbox{tr}\,\Big(C(R)^2\Big)\quad\nonumber\\
&&\qquad\qquad\qquad\qquad\qquad\quad  +\, \mbox{tr}\,C(R) \Big(-\frac{3}{2}C_2 + N_f T(R) \Big) \Big(-\frac{5}{2} + \frac{3}{2}\zeta(3)\Big) \Big]\Big\} + O(\alpha_s^3).\qquad
\end{eqnarray}

For the particular case of the group $G=SU(N)$ and the matter superfields in the (anti)fundamental representation the result (\ref{D_Original_Three_Loop}) is written in Appendix \ref{Appendix_Different_Notation}.

Let us compare the expression (\ref{D_Original_Three_Loop}) with the two-loop anomalous dimension in $\overline{\mbox{DR}}$ scheme. It can be found, e.g., in Ref. \cite{Jack:1996vg}, and, in the notation of this paper, is written as

\begin{equation}\label{Anomalous_Dimension}
\widetilde\gamma(\alpha_s)_i{}^j = -\frac{\alpha_s}{\pi} C(R)_i{}^j + \frac{\alpha_s^2}{2\pi^2}\Big[\Big(-\frac{3}{2} C_2 + N_f T(R)\Big) C(R)_i{}^j + \left(C(R)^2\right)_i{}^j\Big] + O(\alpha_s^3).
\end{equation}

\noindent
We see that the NSVZ-like equation (\ref{NSVZ_Like_Relation}) is not satisfied,

\begin{eqnarray}\label{NSVZ_Like_Relation_HD_Three_Loop}
\bm{D}(\alpha_s) = \frac{3}{2} \sum_{\alpha=1}^{N_f}q_{\alpha}^2\, \Big\{\Big(\mbox{dim}(R) - \mbox{tr}\, \widetilde\gamma(\alpha_{s}) \Big)
+ \frac{\alpha_s^2}{\pi^2}\, \beta_0\, \mbox{tr}\, C(R)\Big(-2+\frac{3}{2}\zeta(3)\Big)\Big\}  + O(\alpha_s^3).
\end{eqnarray}

\noindent
In this equation $\beta_0$ denotes the first coefficient of the SQCD $\beta$-function\footnote{Following Ref. \cite{Kataev:2013eta} in this paper we mark the $\beta$-function (standardly) defined in terms of the renormalized couplings by a tilde.} (such that $\widetilde\beta(\alpha_s) = \alpha_s^2 \beta_0/\pi + O(\alpha_s^3)$), which is given by the expression

\begin{equation}\label{Beta0}
\beta_0 = - \frac{3}{2}C_2 + N_f T(R).
\end{equation}

\noindent
(At present the SQCD $\beta$-function is known up to the three-loop approximation \cite{Harlander:2009mn}.)

Below in Sect. \ref{Section_NSVZ} we will see that it is possible to tune a finite renormalization of the matter superfields in such a way that the NSVZ-like relation will be valid for $\bm{D}(\alpha_s)$, while finite renormalizations of the SQCD coupling constant do not affect the NSVZ-like equation in the considered order.

\section{The three-loop $D$-function in the $\overline{\mbox{DR}}$ scheme for all renormalization constants}
\hspace{\parindent}\label{Section_Adler_Function_DR}

Calculating the three-loop result (\ref{D_Original_Three_Loop}) for the function $\bm{D}(\alpha_s)$ defined by Eq. (\ref{D_Function_Original}) we use the $\overline{\mbox{DR}}$ prescription only for the renormalization of the SQCD coupling constant and the matter superfields, while for the SQCD-renormalized electromagnetic coupling constant we use the MOM-like prescription (\ref{MOM_Scheme}). Now, let us calculate the three-loop result for the function (\ref{D_Function_Renormalized}) in the case when the $\overline{\mbox{DR}}$ scheme is used for both coupling constants. As we will see, the corresponding perturbative expression is different from the one for the Adler function defined by Eq. (\ref{D_Function_Original}), but is related to it by a finite renormalization. The result can be obtained from Eq. (\ref{D_Function_HD}) for a certain values of the finite constants $b_{11}$, $b_{12}$, and $d_2$. As for $b_{11}$ and $b_{12}$, they determine the subtraction scheme for the SQCD coupling constant and, therefore, as in the case considered in the previous section, are given by Eq. (\ref{B_Values}). Thus, it is necessary to find only the value of the constant $d_2$. This can be done by comparing the expressions for the functions $d^{-1}$ calculated with the higher covariant derivative regularization and with the dimensional reduction in the $\overline{\mbox{DR}}$ scheme. The $\overline{\mbox{DR}}$ result can be obtained similarly to the case of ${\cal N}=1$ SQED considered in \cite{Aleshin:2015qqc},

\begin{equation}\label{DInverse_Function_DR}
d^{-1}_{\overline{\mbox{\scriptsize DR}}}=\alpha^{-1} + \frac{1}{\pi}\sum_{\alpha=1}^{N_{f}}q_{\alpha}^2\, \mbox{dim}(R)\Big(\ln\frac{\mu}{P} + 1\Big)
+\frac{\alpha_s}{\pi^2}\sum_{\alpha=1}^{N_{f}}q_{\alpha}^2\, \mbox{tr}\,C(R) \Big(\ln\frac{\mu}{P}+\frac{7}{4}-\frac{3}{2}\zeta(3)\Big) + O(\alpha_s^2).
\end{equation}

\noindent
Taking into account that the renormalized all-loop expression for Green function should not depend on the regularization and renormalization prescriptions, we should equate the results (\ref{DInverse_Function_HD}) and (\ref{DInverse_Function_DR}),

\begin{equation}\label{Renormalized_D_Equality}
d^{-1}_{\mbox{\scriptsize HD}} = d^{-1}_{\overline{\mbox{\scriptsize DR}}}.
\end{equation}

\noindent
From this equation we find the values of the constants $d_1$ and $d_2$ corresponding to the $\overline{\mbox{DR}}$ scheme,

\begin{equation}\label{D_Values_DR}
d_1 = \ln a;\qquad d_2 = -1/4.
\end{equation}

\noindent
Note that again they have been derived from the equality of the finite parts of the Green functions. Substituting the values of the finite constants (\ref{B_Values}) and (\ref{D_Values_DR}) into Eq. (\ref{D_Function_HD}) we obtain the result for the $D$-function (\ref{D_Function_Renormalized}) in the $\overline{\mbox{DR}}$ scheme

\begin{eqnarray}\label{D_Function_DR}
&& \widetilde D(\alpha_s)=
\frac{3}{2}\sum_{\alpha=1}^{N_{f}}q_{\alpha}^2\, \Big\{\mbox{dim}(R)+\frac{\alpha_s}{\pi}\mbox{tr}\,C(R)+\frac{\alpha^2_s}{\pi^2}
\Big[-\frac{1}{2}\mbox{tr}\,(C(R)^2) \qquad\nonumber\\
&&\qquad\qquad\qquad\qquad\qquad\qquad - \frac{3}{4} \mbox{tr}\,C(R) \Big(-\frac{3}{2} C_2 + N_{f} T(R)\Big)\Big]\Big\} + O(\alpha_s^3).\qquad
\end{eqnarray}

We note that in the Abelian case ($G\to U(1)$) it is related to the three-loop function $\widetilde\beta(\alpha)/\alpha^2$ in ${\cal N}=1$ SQED with $N_f$ flavors which can be found from the results of Ref. \cite{Jack:1996vg}. In this case we really reproduce this function setting $\mbox{dim}(R) \to 1$, $C(R)_i{}^j \to 1$, $T(R)\to 1$, $C_2 \to 0$ and making the replacement $3\pi/2 \sum_{\alpha=1}^{N_f} q_\alpha^2 \to N_f$. This fact can be considered as a confirmation of the calculation correctness.

Also we see that in the $\overline{\mbox{DR}}$ scheme the $D$-function does not satisfy the NSVZ-like relation (\ref{NSVZ_Like_Relation}),

\begin{equation}\label{NSVZ_Like_Relation_DR}
\widetilde D(\alpha_s)=\frac{3}{2}\sum_{\alpha=1}^{N_{f}}q_{\alpha}^2\,\Big\{\Big(\mbox{dim}(R)-\mbox{tr}\,\widetilde\gamma(\alpha_s)\Big) - \frac{\alpha_s^2}{4\pi^2}\, \beta_0\,\mbox{tr}\,C(R)\Big\} + O(\alpha_s^3),
\end{equation}

\noindent
where the anomalous dimension in the $\overline{\mbox{DR}}$ scheme is given by Eq. (\ref{Anomalous_Dimension}).

Starting from Eq. (\ref{D_Function_DR}) it is possible to calculate the $D$-function (\ref{D_Function_Bare}) defined in terms of the bare couplings for the theory regularized by the dimensional reduction. The details of this calculation are described in Appendix \ref{Appendix_Bare_D}, and the result is

\begin{eqnarray}\label{D_Function_Bare_Result}
&& D(\alpha_{s0})=\frac{3}{2}\sum_{\alpha=1}^{N_f} q_{\alpha}^2\,\Big\{\mbox{dim}(R)+\frac{\alpha_{s0}}{\pi}\mbox{tr}\,C(R) + \frac{\alpha_{s0}^2}{\pi^2}\Big[-\frac{1}{2}\mbox{tr}\, \left(C(R)^2\right)\nonumber\\
&&\qquad\qquad\qquad\qquad\qquad\qquad -\Big(\frac{3}{4}+\frac{1}{2\varepsilon}\Big)\, \mbox{tr}\,C(R)\, \Big(-\frac{3}{2} C_2 + N_f T(R) \Big)\Big]\Big\} + O(\alpha_{s0}^3).\qquad
\end{eqnarray}

\noindent
This expression (for a fixed regularization) is independent of the renormalization prescription, see, e.g. Ref. \cite{Kataev:2013eta}. Therefore, it does not contain any finite constants defining the subtraction scheme
(i.e. Eq. (\ref{D_Function_Bare_Result}) is valid for an arbitrary renomalization procedure in the case of using the regularization by dimensional reduction). It is worth mentioning that the result (\ref{D_Function_Bare_Result}) differs from the $D$-function (\ref{D_Function_DR}) calculated in the $\overline{\mbox{DR}}$ scheme by the presence of the $\varepsilon$-pole, as it should be \cite{Aleshin:2016rrr}. Also we see that

\begin{equation}
D(\alpha_{s0})=\frac{3}{2}\sum_{\alpha=1}^{N_{f}}q_{\alpha}^2\,\Big\{\Big(\mbox{dim}(R)-\mbox{tr}\,\gamma(\alpha_{s0})\Big) - \frac{\alpha_{s0}^2}{\pi^2}\,\beta_0\, \mbox{tr}\,C(R)
\Big(\frac{1}{4}+\frac{1}{2\varepsilon}\Big)\Big\} + O(\alpha_{s0}^3),
\end{equation}

\noindent
where the anomalous dimension $\gamma(\alpha_{s0})$ defined in terms of the bare coupling constant for the theory regularized by the dimensional reduction has the form

\begin{equation}
\gamma(\alpha_{s0})_i{}^j = -\frac{\alpha_{s0}}{\pi} C(R)_i{}^j + \frac{\alpha_{s0}^2}{2\pi^2}\Big[ \Big(-\frac{3}{2} C_2 + N_f T(R)\Big) C(R)_i{}^j + \left(C(R)^2\right)_i{}^j\Big] + O(\alpha_{s0}^3),
\end{equation}

\noindent
see Appendix \ref{Appendix_Bare_D}.  Thus, unlike the higher covariant derivative regularization \cite{Shifman:2014cya,Shifman:2015doa}, the three-loop Adler D-function defined in terms of the bare coupling constant  calculated with the dimensional reduction does not satisfy the NSVZ-like relation and explicitly depends on $\varepsilon$ similar to the ${\cal N}=1$ SQED case considered in \cite{Aleshin:2016rrr}.

\section{Scheme dependence of the $D$-function}
\hspace*{\parindent}\label{Section_Scheme_Dependence}

In the previous sections we have demonstrated that the NSVZ-like equation for the $D$-function is not satisfied in the $\overline{\mbox{DR}}$ scheme. This is a typical situation, see, e.g., \cite{Jack:1996vg,Jack:1996cn,Jack:1998uj,Harlander:2006xq,Harlander:2009mn,Mihaila:2013wma}. However, the NSVZ-like relations can be restored by a properly tuned finite renormalization

\begin{equation}\label{Finite_Renormalization_Original}
\frac{1}{\alpha} \to \frac{1}{\alpha'} =\frac{1}{\alpha} + f(\alpha_s); \qquad \alpha_s \to \alpha_s'(\alpha_s);\qquad Z'(\alpha'_s)_i{}^j = z(\alpha_s)_i{}^k Z(\alpha_s)_k{}^j.
\end{equation}

\noindent
Under this transformation the $D$-function (defined in terms of the renormalized coupling constant) changes as

\begin{equation}\label{D_Transformation}
\widetilde D'(\alpha_s') = -\frac{3\pi}{2}\frac{d}{d\ln\mu}\Big(\frac{1}{\alpha'}\Big) = \widetilde D(\alpha_s) -\frac{3\pi}{2}\, \frac{df(\alpha_s)}{d\alpha_s}\, \widetilde\beta(\alpha_s),
\end{equation}

\noindent
where it is assumed that the right hand side is expressed in terms of $\alpha_s'$. In the lowest order the finite renormalization (\ref{Finite_Renormalization_Original}) can be presented in the form

\begin{eqnarray}\label{Alpha_Change}
&& \frac{1}{\alpha'} = \frac{1}{\alpha} + \frac{1}{\pi} \sum\limits_{\alpha=1}^{N_f} q_\alpha^2\, \Big(f_0 + f_1 \frac{\alpha_s}{\pi} + O(\alpha_s^2)\Big);\\
\label{AlphaS_Change}
&& \frac{1}{\alpha_s'} = \frac{1}{\alpha_s} + \frac{\delta_0}{\pi} + O(\alpha_s);\vphantom{\Bigg(}\\
\label{Z_Change}
&& z(\alpha_s)_i{}^j = \delta_i^j + \frac{\alpha_s}{\pi}\,(z_1)_i{}^j + O(\alpha_s^2),\vphantom{\Bigg(}
\end{eqnarray}

\noindent
where $f_0$, $f_1$, $\delta_0$, and $(z_1)_i{}^j$ are finite constants. In particular, this implies that in the considered approximation $\alpha_s = \alpha_s' + \delta_0\, (\alpha_s')^2/\pi + O\left((\alpha_s')^3\right)$.
Note that it is reasonable to include into the finite constants some group theory factors. Really, the SQCD quantum corrections to $1/\alpha$ are proportional to $\mbox{dim}(R)$ and $\mbox{tr}\, C(R)$ in the one- and two-loop approximations, respectively. The one-loop quantum corrections to $1/\alpha_s$ are proportional to either $C_2$ or $N_f T(R)$, and the one-loop quantum correction to the matter renormalization contains $C(R)_i{}^j$. That is why it is expedient to consider only the finite renormalizations with

\begin{eqnarray}\label{Modified_F}
&& f_0 = \mbox{dim}(R)\, \widetilde f_0;\qquad f_1 = \mbox{tr}\, C(R)\, \widetilde f_1;\vphantom{\Big(}\\
\label{Modified_Delta}
&& \delta_0 = - \frac{3}{2} C_2\, \widetilde \delta_{01} + N_f T(R)\, \widetilde \delta_{02};\vphantom{\Big(}\\
\label{Modified_Z}
&& (z_1)_i{}^j = C(R)_i{}^j\, \widetilde z_1,\vphantom{\Big(}
\end{eqnarray}

\noindent
where the factor $-3/2$ in Eq. (\ref{Modified_Delta}) is introduced for the further convenience. Substituting Eqs. (\ref{Alpha_Change}) --- (\ref{Z_Change}) into Eq. (\ref{D_Transformation}), we obtain that the $D$-function in the new subtraction scheme is given by the expression

\begin{eqnarray}\label{D_Tranformed}
&& \widetilde D'(\alpha_s') = \widetilde D(\alpha_s') + \frac{(\alpha_s')^2}{\pi}\,\delta_0\, \frac{d\widetilde D(\alpha_s')}{d\alpha_s'} -\frac{3\pi}{2} \frac{df(\alpha_s)}{d\alpha_s} \beta_0 \frac{\alpha_s^2}{\pi} + O\left((\alpha_s')^3\right)\nonumber\\
&&\qquad\qquad\qquad\qquad = \widetilde D(\alpha_s') + \frac{3}{2}\sum\limits_{\alpha=1}^{N_f} q_\alpha^2\, \frac{(\alpha_s')^2}{\pi^2}\, \Big(\delta_0\, \mbox{tr}\, C(R) - f_1 \beta_0 \Big)  + O\left((\alpha_s')^3\right).\qquad
\end{eqnarray}

The equation describing the transformation of the anomalous dimension under a finite renormalization is well known \cite{Vladimirov:1979my}. For the finite renormalizations (\ref{Alpha_Change}) --- (\ref{Z_Change}) it gives

\begin{eqnarray}\label{Gamma_Transformed}
&& \widetilde\gamma'(\alpha_s')_i{}^j = \widetilde\gamma(\alpha_s)_i{}^j + \frac{d\ln z_i{}^j}{d\alpha_s}\, \widetilde\beta(\alpha_s)\nonumber\\
&& \qquad\quad = \widetilde\gamma(\alpha_s')_i{}^j + \frac{(\alpha_s')^2}{\pi}\, \delta_0\, \frac{d\gamma(\alpha_s')_i{}^j}{d\alpha_s'}  + \frac{(\alpha_s')^2}{\pi}\,\beta_0\, \frac{d\ln z_i{}^j}{d\alpha_s}  + O\left((\alpha_s')^3\right)\nonumber\\
&&\qquad\qquad\qquad = \widetilde\gamma(\alpha_s')_i{}^j + \frac{(\alpha_s')^2}{\pi^2}\,\Big(\beta_0 (z_1)_i{}^j -\delta_0\, C(R)_i{}^j\Big) + O\left((\alpha_s')^3\right).\qquad
\end{eqnarray}

From the equations presented above it is possible to find out how various coefficients of the perturbative expansions for the $D$-function and the anomalous dimension transform under the finite renormalization and which of them are scheme independent. For this purpose we take the finite constants in the form (\ref{Modified_F}) --- (\ref{Modified_Z}). Then (after the formal replacement $\alpha_s'\to \alpha_s$) we obtain

\begin{eqnarray}\label{New_D}
&&\hspace*{-5mm} \widetilde D'(\alpha_s) = \widetilde D(\alpha_s) + \frac{3}{2}\sum\limits_{\alpha=1}^{N_f} q_\alpha^2\, \frac{\alpha_s^2}{\pi^2}\, \Big(-\frac{3}{2}C_2\, \widetilde \delta_{01} + N_f T(R)\, \widetilde\delta_{02}\,  - \widetilde f_1 \beta_0 \Big)\, \mbox{tr}\, C(R) + O(\alpha_s^3);\qquad\\
\label{New_Gamma}
&&\hspace*{-5mm} \widetilde\gamma'(\alpha_s)_i{}^j = \widetilde\gamma(\alpha_s)_i{}^j + \frac{\alpha_s^2}{\pi^2}\,\Big(\beta_0 \widetilde z_1 + \frac{3}{2}C_2\, \widetilde \delta_{01} - N_f T(R)\, \widetilde \delta_{02} \Big)\, C(R)_i{}^j + O(\alpha_s^3).
\end{eqnarray}

\noindent
The consequences of these equations will be discussed below in Sect. \ref{Section_Beta_Expansion} in more detail.

\section{The NSVZ-like scheme with dimensional reduction in the three-loop approximation}
\hspace*{\parindent}\label{Section_NSVZ}

Using a possibility of making finite renormalizations it is possible to tune a substraction scheme in such a way to restore the NSVZ-like relation (\ref{NSVZ_Like_Relation}). Using the transformations of the $D$-function and the anomalous dimension given by Eqs. (\ref{D_Tranformed}), (\ref{Gamma_Transformed}), (\ref{New_D}), and (\ref{New_Gamma}) we see that the difference between the left and right hand sides of Eq. (\ref{NSVZ_Like_Relation}) changes as

\begin{eqnarray}\label{NSVZ_Finite_Renormalization}
&& \left[\widetilde D'(\alpha_s) - \frac{3}{2} \smash{\sum\limits_{\alpha=1}^{N_f}} q_\alpha^2 \Big(\mbox{dim}(R)- \mbox{tr}\, \widetilde\gamma'(\alpha_s)\Big)\right] -\left[ \widetilde D(\alpha_s) - \frac{3}{2} \smash{\sum\limits_{\alpha=1}^{N_f}} q_\alpha^2 \Big(\mbox{dim}(R)- \mbox{tr}\, \widetilde\gamma(\alpha_s)\Big)\right]\quad\nonumber\\
&& = \frac{3}{2}\sum\limits_{\alpha=1}^{N_f} q_\alpha^2\, \frac{\alpha_s^2}{\pi^2}\, \beta_0\, \Big(\mbox{tr}\, z_1 -f_1\Big) + O(\alpha_s^3) = \frac{3}{2}\sum\limits_{\alpha=1}^{N_f} q_\alpha^2\, \frac{\alpha_s^2}{\pi^2}\, \beta_0\, \Big(\widetilde z_1 - \widetilde f_1\Big)\, \mbox{tr}\, C(R) + O(\alpha_s^3).\vphantom{\frac{1}{2}}\qquad
\end{eqnarray}

\noindent
From this equation we see that in the considered approximation the NSVZ-like relation can be broken only by terms proportional to the first coefficient of the ${\cal N}=1$ SQCD $\beta$-function.

The constant $\delta_0$ (or, equivalently, the constants $\widetilde \delta_{01}$ and $\widetilde \delta_{02}$) does not enter this equation. This implies that it is impossible to restore Eq. (\ref{NSVZ_Like_Relation}) by making a finite renormalization of the SQCD coupling constant only. From Eq. (\ref{NSVZ_Finite_Renormalization}) it is clear that for this purpose we need either to change the SQCD-renormalized electromagnetic coupling constant or to perform the finite renormalization of the matter superfields.

Now, let us construct finite renormalizations restoring the NSVZ scheme in the case of using the various renormalization prescriptions described above with the help of Eq. (\ref{NSVZ_Finite_Renormalization}).

First, we consider the Adler function $\bm{D}(\alpha_s)$ defined by Eq. (\ref{D_Function_Original}) calculated in the $\overline{\mbox{DR}}$-scheme. Certainly, this prescription should be used for the renormalization of the SQCD coupling constant and the matter superfields only, because the SQCD-renormalized electromagnetic coupling constant is unambiguously determined by Eq. (\ref{MOM_Scheme}). Therefore, there is no arbitrariness in choosing the function $f(\alpha_s)$ in Eq. (\ref{Finite_Renormalization_Original}). This implies that in this case we should set $f_0=0$, $f_1=0$, so that the only possibility to restore the NSVZ-like equation is to make the finite renormalization with

\begin{equation}\label{D_Original_Finite_Renormalization}
(z_1)_i{}^j = \Big(2-\frac{3}{2}\zeta(3)\Big)\, C(R)_i{}^j \qquad\mbox{or,\ equivalently,}\qquad \widetilde z_1 = 2-\frac{3}{2}\zeta(3),
\end{equation}

\noindent
while the parameter $\delta_0$ can take arbitrary values.\footnote{Note that the parameter $\delta_0$ does not also affect the NSVZ relation for the ${\cal N}=1$ SQCD $\beta$-function. According to \cite{Jack:1996vg}, in the three-loop approximation the NSVZ scheme for ${\cal N}=1$ supersymmetric gauge theories regularized by dimensional reduction is constructed by tuning the next coefficient $\delta_1$ defined by the equation $1/\alpha_s' = 1/\alpha_s + \delta_0/\pi + \delta_1 \alpha_s/\pi^2 + O(\alpha_s^2)$.} This finite renormalization removes the last term in Eq. (\ref{NSVZ_Like_Relation_HD_Three_Loop}), so that Eq. (\ref{NSVZ_Like_Relation}) will be satisfied by the functions $\bm{D}'(\alpha_s')$ and $\widetilde\gamma'(\alpha_s')$. Note that the renormalization scheme which is obtained after the transformation (\ref{D_Original_Finite_Renormalization}) can be also defined with the help of the boundary condition imposed on the SQCD coupling constant and the renormalization constant of the matter superfields $Z_i{}^j$,

\begin{eqnarray}
&& \alpha_{s}^{-1}\Big(\alpha_{s0},\,\ln\frac{\bar\Lambda}{\mu}=0,\,\frac{1}{\varepsilon}\to 0\Big) = \alpha_{s0}^{-1} +  \frac{\delta_0}{\pi} + O(\alpha_{s0});\nonumber\\
&& Z_i{}^j\Big(\alpha_s,\,\ln\frac{\bar\Lambda}{\mu}=0,\,\frac{1}{\varepsilon}\to 0\Big) = \delta_i^j + \frac{\alpha_{s}}{\pi}\Big(2-\frac{3}{2}\zeta(3)\Big)\, C(R)_i{}^j + O(\alpha_{s}^2),
\end{eqnarray}

\noindent
where the (formal) condition $1/\varepsilon \to 0$ removes $\varepsilon$-poles. If we choose

\begin{equation}\label{Delta0_Equality}
\widetilde\delta_{01} = \widetilde\delta_{02} \equiv \widetilde\delta_0,
\end{equation}

\noindent
then in this renormalization scheme the Adler function and the anomalous dimension take the form

\begin{eqnarray}
&& \bm{D}(\alpha_s)=
\frac{3}{2}\sum_{\alpha=1}^{N_{f}}q_{\alpha}^2\,\Big\{\mbox{dim}(R)+\frac{\alpha_s}{\pi}\mbox{tr}\,C(R)+\frac{\alpha^2_s}{\pi^2}
\Big[\, - \frac{1}{2}\mbox{tr}\,\Big(C(R)^2\Big)\quad\nonumber\\
&&\qquad\qquad\qquad\qquad\qquad\qquad\qquad\qquad\  + \beta_0\, \mbox{tr}\,C(R) \Big(-\frac{5}{2} + \frac{3}{2}\zeta(3) + \widetilde \delta_0\Big) \Big]\Big\} + O(\alpha_s^3);\\
&& \widetilde\gamma(\alpha_s)_i{}^j = -\frac{\alpha_s}{\pi} C(R)_i{}^j + \frac{\alpha_s^2}{\pi^2}\,  \Big[\beta_0 \Big(\frac{5}{2} - \frac{3}{2}\zeta(3) - \widetilde \delta_0\Big) C(R)_i{}^j  +\frac{1}{2}\left(C(R)^2\right)_i{}^j\Big] + O(\alpha_s^3)\qquad\quad
\end{eqnarray}

\noindent
and evidently satisfy the NSVZ-like relation in the considered approximation.

Next, we construct the NSVZ-like scheme for the function $\widetilde D(\alpha_s)$ defined by Eq. (\ref{D_Function_Renormalized}) calculated in the $\overline{\mbox{DR}}$ scheme for all renormalization constants. In this case it is possible to make finite renormalization of both coupling constants and of the matter superfields. From Eq. (\ref{NSVZ_Like_Relation_DR}) we see that the parameters in Eqs. (\ref{Alpha_Change}) --- (\ref{Z_Change}) and (\ref{Modified_F}) --- (\ref{Modified_Z}) should satisfy the condition

\begin{equation}
\widetilde z_1 - \widetilde f_1 = \frac{1}{4}.
\end{equation}

\noindent
The corresponding renormalization prescription can be formulated by imposing the boundary conditions

\begin{eqnarray}\label{NSVZ_Boundary_Conditions}
&& \alpha^{-1}\Big(\alpha_0,\alpha_{s0},\,\ln\frac{\bar\Lambda}{\mu}=0,\,\frac{1}{\varepsilon}\to 0\Big) = \alpha_0^{-1} + \frac{1}{\pi} \smash{\sum\limits_{\alpha=1}^{N_f}} q_\alpha^2 \Big(\mbox{dim}(R) \widetilde f_0 + \mbox{tr}\, C(R)\, \widetilde f_1 \frac{\alpha_{s0}}{\pi} + O\big(\alpha_{s0}^2\big) \Big);\qquad\nonumber\\
&& \alpha_s^{-1}\Big(\alpha_{s0},\,\ln\frac{\bar\Lambda}{\mu}=0,\,\frac{1}{\varepsilon}\to 0\Big) = \alpha_{s0}^{-1} +  \frac{\delta_0}{\pi} + O(\alpha_{s0});\nonumber\\
&& Z_i{}^j\Big(\alpha_s,\,\ln\frac{\bar\Lambda}{\mu}=0,\,\frac{1}{\varepsilon}\to 0\Big) = \delta_i^j + \frac{\alpha_s}{\pi}\Big(\frac{1}{4} + \widetilde f_1\Big)\, C(R)_i{}^j + O(\alpha_s^2).
\end{eqnarray}

\noindent
We see that (similar to the case of ${\cal N}=1$ SQED considered in \cite{Goriachuk:2018cac}) the NSVZ scheme is not unique and the conditions (\ref{NSVZ_Boundary_Conditions}) describe the class of the NSVZ schemes parameterized by the finite constants $\delta_0$, $\widetilde f_0$, and $\widetilde f_1$. If we choose $\widetilde\delta_{01} = \widetilde\delta_{02}$, so that $\delta_0 = \beta_0 \widetilde \delta_0$, then in this renormalization scheme the $D$-function and the anomalous dimension are given by the expressions

\begin{eqnarray}\label{New_D_Short}
&& \widetilde D(\alpha_s)=
\frac{3}{2}\sum_{\alpha=1}^{N_{f}}q_{\alpha}^2\,\Big\{\mbox{dim}(R)+\frac{\alpha_s}{\pi}\mbox{tr}\,C(R)+\frac{\alpha^2_s}{\pi^2}
\Big[\, - \frac{1}{2}\mbox{tr}\,\Big(C(R)^2\Big)\quad\nonumber\\
&&\qquad\qquad\qquad\qquad\qquad\qquad\qquad\qquad\,  + \beta_0\, \mbox{tr}\,C(R) \Big( - \frac{3}{4} +\widetilde \delta_0 -\widetilde f_1\Big) \Big]\Big\} + O(\alpha_s^3);\\
\label{New_Gamma_Short}
&& \widetilde\gamma(\alpha_s)_i{}^j = -\frac{\alpha_s}{\pi} C(R)_i{}^j + \frac{\alpha_s^2}{\pi^2}\,  \Big[\beta_0 \Big(\frac{3}{4} - \widetilde \delta_0 + \widetilde f_1\Big) C(R)_i{}^j  +\frac{1}{2}\left(C(R)^2\right)_i{}^j\Big] + O(\alpha_s^3).\qquad
\end{eqnarray}

Finally, we note that for the function $D(\alpha_{s0})$ defined by Eq. (\ref{D_Function_Bare}), it is certainly impossible to restore the NSVZ-like equation by making any finite renormalization, because $D(\alpha_{s0})$ is scheme-independent for a fixed regularization.

\section{$\beta_0$-expansion}
\hspace*{\parindent}\label{Section_Beta_Expansion}

It is convenient to present the expression for the Adler $D$-function using the $\beta$-expansion formalism proposed in Ref. \cite{Mikhailov:2004iq}. According to this formalism, the scheme-dependent coefficients of the massless perturbative quantum corrections are expanded in the monomials of the QCD $\beta$-function coefficients. The remaining terms are not affected by the renormalization prescription for the QCD coupling constant and satisfy certain relations, which can be obtained in the so-called conformal symmetry limit \cite{Kataev:2010du,Kataev:2013vua}. The $\beta$-expansion formalism generalizes the Brodsky-Lepage-Mackenzie (BLM) scale-fixing
prescription \cite{Brodsky:1982gc} (see Refs. \cite{Mikhailov:2004iq, Brodsky:2013vpa,Kataev:2014jba,Ma:2015dxa,Kataev:2016aib} for the detailed discussion).

There are various procedures for fixing the coefficients of $\beta$-expanded expressions for the Adler $D$-function \cite{Mikhailov:2004iq,Kataev:2010du,Brodsky:2013vpa,Kataev:2014jba,Kataev:2016aib,
Cvetic:2016rot}. Here we apply the $\beta$-expansion procedure to the next-to-leading order (NLO) of SQCD expression for the Adler $D$-function. The results can be compared with the ones for QCD, see Refs. \cite{Brodsky:2013vpa,Kataev:2014jba,Ma:2015dxa,Kataev:2016aib,Cvetic:2016rot}. Note that in the considered approximation the $\beta$-expansion coincides with the $\beta_0$-expansion considered in Ref. \cite{Beneke:1994qe}, where $\beta_0$ is the first (scheme-independent) coefficient of the $\beta$-function.

Within the $\beta_0$-expansion the results for the SQCD Adler function and for the anomalous dimension of the matter superfields are presented in the form
\begin{eqnarray}\label{Beta_Expansion_D}
&& D(\alpha_s) = \frac{3}{2}\sum\limits_{\alpha=1}^{N_f} q_{\alpha}^2\, \Big\{\mbox{dim}(R)+ \mbox{tr}\, C(R) \sum\limits_{n=1}^{2} D_n \Big(\frac{\alpha_s}{\pi}\Big)^n\Big\}+O(\alpha_s^3);\qquad\\
\label{Beta_Expansion_Gamma}
&& \widetilde\gamma(\alpha_s)_i{}^j = \sum\limits_{n=1}^{2} \big(\gamma_n\big)_i{}^j \Big(\frac{\alpha_s}{\pi}\Big)^n + O(\alpha_s^3),
\end{eqnarray}
where $D(\alpha_s)$ stands for either $\bm{D}(\alpha_s)$ or $\widetilde D(\alpha_s)$. Similar equations can be also written for $D(\alpha_{s0})$ and $\gamma(\alpha_{s0})_i{}^j$, but then it is necessary to make a formal substitution $\alpha_{s} \to \alpha_{s0}$. The coefficients in Eqs. (\ref{Beta_Expansion_D}) and (\ref{Beta_Expansion_Gamma}) have the structure
\begin{eqnarray}\label{Beta_Expansion_Coefficients}
&& D_1 = D_1[0];\qquad\qquad D_2 = \beta_0\, D_2[1] + D_2[0];\vphantom{\Big(}\\
&& \big(\gamma_1\big)_i{}^j = \gamma_1[0]_i{}^j;\qquad\ \big(\gamma_2\big)_i{}^j = \beta_0\, C(R)_i{}^j\, \gamma_2[1] + \gamma_2[0]_i{}^j,\vphantom{\Big(}
\end{eqnarray}
where $\beta_0$ is defined by Eq. (\ref{Beta0}). For the renormalization prescriptions which do not break the NSVZ-like relation (\ref{NSVZ_Like_Relation}) the coefficients of Eqs. (\ref{Beta_Expansion_D}) and (\ref{Beta_Expansion_Gamma}) are related by the equations
\begin{equation}
D_n\, \mbox{tr}\, C(R)  = - \mbox{tr}\, \gamma_n \qquad \mbox{for}\qquad n=1,2.
\end{equation}

\begin{table}[h]
\small\
\begin{tabular}{|c|c|c|c|c|c|c|}
\hline
Function & Regularization & $\alpha$ & $\alpha_s$ and $Z_i{}^j$ & $D_2[1]\vphantom{\Bigg(}$ & $\gamma_2[1]$ & NSVZ\\
\hline\hline
$\bm{D}(\alpha_s)$ & DRED & MOM-like & $\overline{\mbox{DR}}$ & ${\displaystyle -\frac{5}{2}+ \frac{3}{2}\zeta(3)}\vphantom{\Bigg(}$ & ${\displaystyle \frac{1}{2}\vphantom{\Bigg(}}$ & $-$\\
\hline
$\bm{D}(\alpha_s)$ & DRED & MOM-like & $\mbox{NSVZ}$ & ${\displaystyle -\frac{5}{2}+\frac{3}{2}\zeta(3) + \widetilde\delta_0\vphantom{\Bigg(}}$ & ${\displaystyle \, \frac{5}{2} - \frac{3}{2}\zeta(3) - \widetilde\delta_0\vphantom{\Bigg(}}$ & $+$\\
\hline
$\widetilde{D}(\alpha_s)$ & DRED & $\overline{\mbox{DR}}$ & $\overline{\mbox{DR}}$ & ${\displaystyle -\frac{3}{4}\vphantom{\Bigg(}}$ & ${\displaystyle \frac{1}{2}\vphantom{\Bigg(}}$ & $-$\\
\hline
$\widetilde{D}(\alpha_s)$ & DRED & \parbox{1.5cm}{$\overline{\mbox{DR}}$+finite renormal.} & $\mbox{NSVZ}$ & ${\displaystyle  -\frac{3}{4} + \widetilde\delta_0 - \widetilde f_1}\vphantom{\Bigg(}$ & ${\displaystyle \frac{3}{4} - \widetilde\delta_0 + \widetilde f_1}$ & $+$\\
\hline
$\widetilde{D}(\alpha_s)$ & HD; $a_\varphi = a$ & HD+MSL & HD+MSL & ${\displaystyle -1-\ln a\vphantom{\Bigg(}}$ & ${\displaystyle 1+\ln a\vphantom{\Bigg(}}$ & $+$\\
\hline
$D(\alpha_{s0})$ & DRED & arbitrary & arbitrary & ${\displaystyle -\frac{3}{4} -\frac{1}{2\varepsilon}}$ & ${\displaystyle \frac{1}{2}\vphantom{\Bigg(}}$ & $-$\\
\hline
$D(\alpha_{s0})$ & HD; $a_\varphi = a$ & arbitrary & arbitrary & ${\displaystyle -1-\ln a\vphantom{\Bigg(}}$ & ${\displaystyle 1+\ln a\vphantom{\Bigg(}}$ & $+$\\
\hline
\end{tabular}
\normalsize
\caption{Scheme-dependent coefficients of the $\beta_0$-expansions for different definitions of the $D$-function and of the anomalous dimension of the matter superfields in various renormalization schemes.}\label{Table_Scheme_Dependence}
\end{table}

According to Eq. (\ref{New_D_Short}) the structure (\ref{Beta_Expansion_Coefficients}) is not broken by the finite renormalizations satisfying the condition (\ref{Delta0_Equality}). From Eqs. (\ref{New_D_Short}) and (\ref{New_Gamma_Short}) we also see that the coefficients
\begin{equation}
D_1[0] = 1; \qquad D_2[0] = -\frac{\mbox{tr}(C(R)^2)}{2\,\mbox{tr}\, C(R)}; \qquad \gamma_1[0]_i{}^j = - C(R)_i{}^j;\qquad \gamma_2[0]_i{}^j = \frac{1}{2} \left(C(R)^2\right)_i{}^j
\end{equation}
are scheme-independent, while the coefficients $D_2[1]$ and $\gamma_2[1]$ under the finite renormalization (\ref{Finite_Renormalization_Original}) (in which the parameters are given by Eqs. (\ref{Modified_F}) --- (\ref{Modified_Z})) changes as\footnote{The scheme dependence of the coefficient $D_2[1]$ was first noted in Ref. \cite{Chyla:1995ej}.}
\begin{equation}
D_2[1] \to D_2[1] + \widetilde\delta_0 - \widetilde f_1; \qquad \gamma_2[1] \to \gamma_2[1] + \widetilde z_1 - \widetilde \delta_{0}.
\end{equation}
Qualitatively, this can be understood as follows. Let us use for ${\cal N}=1$ SQCD the Banks-Zaks prescription \cite{Banks:1981nn}, when a gauge group and a number of matter supermultiplets are fixed in such a way that $\beta_0=0$. For ${\cal N}=1$ SQCD this implies that $N_f=3C_2/(2T(R))$. Then, involving the argumentation and terminology of Ref. \cite{Kataev:2013vua}, in this case we obtain the conformal symmetry limit, so that the coefficients $D_1[0]$, $D_2[0]$, $\gamma_1[0]_i{}^j$, and $\gamma_2[0]_i{}^j$ surviving in this limit should be scheme independent.

The values of the parameters $D_2[1]$ and $\gamma_2[1]$ for various definitions of the renormalization group functions in various subtraction schemes considered in this paper are collected in Table \ref{Table_Scheme_Dependence}. In particular, for the most interesting case, when the functions $\bm{D}(\alpha_s)$ and $\widetilde\gamma(\alpha_s)_i{}^j$ are calculated in the $\overline{\mbox{DR}}$-scheme (Eqs. (\ref{D_Original_Three_Loop}) and (\ref{Anomalous_Dimension})), they take the form
\begin{equation}\label{D_Result}
D_2[1] = -\frac{5}{2}+ \frac{3}{2}\zeta(3);\qquad \gamma_2[1] = \frac{1}{2}.
\end{equation}
It is necessary to recall that in our notation $\gamma(\alpha_s)_i{}^j$ is the anomalous dimension of the matter superfields which includes quarks and squarks as components, $\gamma_2[1]$ being a coefficient of $\beta_0 C(R)_i{}^j$.
The result (\ref{D_Result}) has a similar structure to the QCD one derived in Refs. \cite{Chetyrkin:1979bj,Dine:1979qh,Chetyrkin:1980sa,Chetyrkin:1980pr,Celmaster:1980ji} in the $O(\alpha_s^2)$ order.

\section{Conclusion}
\hspace*{\parindent}

In this paper we obtain the three-loop expression for the Adler $D$-function (\ref{D_Function_Original}) in the case of using the $\overline{\mbox{DR}}$ scheme for the renormalization of the SQCD coupling constant and the matter superfields. This allows to compare the structures of the perturbation theory series in QCD and SQCD, including the form of the $\beta$-expansion at the $O(\alpha_s^2)$-level, and to demonstrate special features caused by supersymmetry. In particular, supersymmetry leads to the NSVZ-like equation (\ref{NSVZ_Like_Relation}), which is absent in the non-supersymmetric case. This equation allows relating the coefficients $D_n$ and $\gamma_n$ of the $\beta$-expansions for the $D$-function and the anomalous dimension of the matter superfields, respectively.  Although the result (\ref{D_Original_Three_Loop}) does not satisfy the NSVZ-like equation, it is possible to find such a finite renormalization of the matter superfields after that this equation will take place in both original and $\beta$-expanded forms. It is worth mentioning that the NSVZ-like relation cannot be restored by making the finite renormalization of the SQCD coupling constant only, see Eq. (\ref{NSVZ_Finite_Renormalization}).

We have also calculated the $D$-function defined by Eq. (\ref{D_Function_Renormalized}) in the case when  the $\overline{\mbox{DR}}$ scheme is used for constructing all renormalization constants, including the one for the SQCD-renormalized electromagnetic coupling constant. Again, the NSVZ-like equation is not valid in this case, but it can be restored by a finite renormalization of the matter superfields or (and) the SQCD-renormalized electromagnetic coupling constant. The $D$-function (\ref{D_Function_Bare}) defined in terms of the bare couplings does not satisfy the NSVZ-like equation (\ref{NSVZ_Like_Relation}) if the theory is regularized by the dimensional reduction and explicitly depends on $\varepsilon = 4-d$.

The results for the Adler $D$-function obtained in this paper may be used for analyzing some indirect manifestations of supersymmetry, which could be revealed by comparing experimental data with the theoretical predictions. Certainly, for this purpose it will be also necessary to take into account superpartner thresholds appearing due to supersymmetry breaking. Although at present the effects of higher order quantum corrections seem to be too small, they may be useful for future research.

\section*{Acknowledgements}
\hspace*{\parindent}

The authors are very grateful to A.E.Kazantsev for valuable discussions.

The work of S.A. was performed at Institute for Information Transmission Problems. The work of A.K. was supported by the Foundation for the Advancement of Theoretical Physics and Mathematics ``BASIS'' (Grant No. 17-11-120).

\appendix

\section{The relation between various definitions of the $D$-function}
\hspace*{\parindent}\label{Appendix_D_Relation}

Let us demonstrate that the Adler function (\ref{D_Function_Original}) coincides with the $D$-function (\ref{D_Function_Renormalized}) in the case of using the boundary condition (\ref{MOM_Scheme}) for the SQCD renormalized electormagnetic coupling constant. The equation (\ref{MOM_Scheme}) is written in the form

\begin{equation}\label{Pi_Boundary_Condition}
\Pi(\alpha_s,P/\mu=1) = 0,
\end{equation}

\noindent
where $\Pi$ denotes the renormalized polarization operator. With the help of the chain rule for the derivative $d/d\ln\mu$ the $D$-function defined by Eq. (\ref{D_Function_Original}) can be written as

\begin{eqnarray}
&& \bm{D}(\alpha_s) = - 6\pi^2 \frac{\partial}{\partial\ln P} \Pi(\alpha_s,P/\mu) = 6\pi^2 \frac{\partial}{\partial\ln\mu}\Pi(\alpha_s,P/\mu)\nonumber\\
&&\qquad\qquad\qquad\qquad = 6\pi^2 \left(\frac{d}{d\ln\mu}\Pi(\alpha_s,P/\mu)\Big|_{\alpha_{s0}=\mbox{\scriptsize const}} - \widetilde\beta(\alpha_s) \frac{\partial}{\partial \alpha_s}\Pi(\alpha_s,P/\mu)\right).\qquad
\end{eqnarray}

\noindent
(This equation was also written in some earlier papers, see, e.g., \cite{Dine:1979qh,Baikov:2012zm}.) Setting in this equation $P=\mu$ and using the boundary condition (\ref{Pi_Boundary_Condition}) we obtain

\begin{eqnarray}
&&\hspace*{-8mm} \bm{D}(\alpha_s) = 6\pi^2 \frac{d}{d\ln\mu}\Pi(\alpha_s,P/\mu)\Big|_{\alpha_{s0}=\mbox{\scriptsize const}} = \frac{3\pi}{2} \frac{d}{d\ln\mu}\Big(d^{-1}(\alpha,\alpha_s,P/\mu)-\alpha^{-1}\Big)\Big|_{\alpha_0,\alpha_{s0}=\mbox{\scriptsize const};\,\alpha\to 0}\nonumber\\
&&\hspace*{-8mm} = -\frac{3\pi}{2} \frac{d}{d\ln\mu} \alpha^{-1}\Big|_{\alpha_0,\alpha_{s0}=\mbox{\scriptsize const};\,\alpha\to 0} = \widetilde D(\alpha_s).\qquad
\end{eqnarray}

\section{The Adler function $\bm{D}(\alpha_s)$ in the $\overline{\mbox{DR}}$ scheme for $G=SU(N)$ and matter in the (anti)fundamental representation}
\hspace*{\parindent}\label{Appendix_Different_Notation}

Here we will consider the special case of the $SU(N)\times U(1)$ gauge group and the matter superfields lying in the fundamental ($\phi$) and antifundamental ($\widetilde{\phi}$) representations. Moreover, we rewrite the result in another notation system adopted in Ref. \cite{Kataev:2013vua}. If the matter superfields lie in an irreducible representation of the group $G$, then this notation system is related to the one accepted in the present paper as

\begin{eqnarray}
&& C_2 \to C_A;\qquad\ \ \, r = \delta_A^A \to N_A; \qquad\ \ \, T(R) \to T_F;\vphantom{\Big(}\nonumber\\
&& \mbox{dim}(R) =\delta_i^i\to d_F;\qquad C(R)_i{}^j = C(R)\delta_i^j \to C_F \delta_i^j.\vphantom{\Big(}
\end{eqnarray}

\noindent
Then the three-loop result (\ref{D_Original_Three_Loop}) can be presented as

\begin{eqnarray}
&& \bm{D}(\alpha_s)=
\frac{3}{2} d_F \sum_{\alpha=1}^{N_{f}}q_{\alpha}^2\,\Big\{1 + \frac{\alpha_s}{\pi} C_F + \frac{\alpha^2_s}{\pi^2}
\Big[\, - \frac{1}{2} C_F^2 \quad\nonumber\\
&&\qquad\qquad\qquad\qquad\qquad  +\, C_F \Big(-\frac{3}{2}C_A + N_f T_F \Big) \Big(-\frac{5}{2} + \frac{3}{2}\zeta(3)\Big) \Big]\Big\} + O(\alpha_s^3).\qquad
\end{eqnarray}

\noindent
For  the particular case $G = SU(N)$ and the (anti)fundamental representation for the matter

\begin{eqnarray}
C_A=N;\qquad T_F=\frac{1}{2}; \qquad N_A=N^2-1; \qquad d_F=N; \qquad C_F=\frac{N^2-1}{2N}.
\end{eqnarray}

\noindent
Therefore, the function $\bm{D}(\alpha_s)$ for ${\cal N}=1$ SQCD in this case takes the form

\begin{eqnarray}
&& \bm{D}(\alpha_s)= \frac{3}{2} N \sum_{\alpha=1}^{N_{f}}q_{\alpha}^2\,\Big\{1 + \frac{\alpha_s}{\pi}\frac{N^2-1}{2N} + \frac{\alpha^2_s}{\pi^2}
\Big[\, - \frac{1}{2} \Big(\frac{N^2-1}{2N}\Big)^2 \quad\nonumber\\
&&\qquad\qquad\qquad\qquad\qquad\qquad  +\, \frac{N^2-1}{8N} \Big(3N - N_f\Big) \Big(5 - 3\zeta(3)\Big) \Big]\Big\} + O(\alpha_s^3).\qquad
\end{eqnarray}

\noindent
Note that for ${\cal N}=1$ SQCD (unlike the case of usual QCD) the Banks--Zaks prescription $\beta_0=0$ \cite{Banks:1981nn} always gives integer values of $N_f$, namely, $N_f=3N$.

\section{The three-loop $D$-function defined in terms of the bare coupling constant}
\hspace*{\parindent}\label{Appendix_Bare_D}

To calculate the $D$-function defined in terms of the bare coupling constant $\alpha_{s0}$, first, it is necessary to find a relation between the bare and SQCD-renormalized electromagnetic coupling constants. For this purpose we substitute the $\overline{\mbox{DR}}$ result (\ref{D_Function_DR}) into Eq. (\ref{D_Function_Renormalized}), express the SQCD coupling constant $\alpha_s$ in terms of the bare SQCD coupling constant $\alpha_{s0}$ from the $\overline{\mbox{DR}}$ equation

\begin{equation}\label{AlphaS_Relation_DR}
\frac{1}{\alpha_{s0}} = \frac{1}{\alpha_{s}} + \frac{1}{2\pi} \Big(3 C_2 - 2N_f T(R)\Big)\Big(\frac{1}{\varepsilon} + \ln\frac{\bar\Lambda}{\mu}\Big) + O(\alpha_{s}),
\end{equation}

\noindent
and integrate with respect to $\ln\mu$. Certainly, it is necessary to take into account that in the variant of the $\overline{\mbox{DR}}$ scheme adopted in this paper all renormalization constants contain only $\varepsilon$-poles and powers of $\ln\bar\Lambda/\mu$, where $\bar\Lambda$ is given by Eq. (\ref{Bar_Lambda}). The above described procedure gives

\begin{eqnarray}\label{Alpha_Relation_Original}
&& \frac{1}{\alpha}=\frac{1}{\alpha_0}+\frac{1}{\pi}\sum_{\alpha}q_{\alpha}^2\Big\{ \dim(R)\, \ln\frac{\bar\Lambda}{\mu} + \frac{\alpha_{s0}}{\pi} \mbox{tr}\, C(R)\, \ln\frac{\bar\Lambda}{\mu}
+ \frac{\alpha_{s0}^2}{\pi^2}\, \Big[-\frac{1}{2}\mbox{tr} \left(C(R)^2\right)\ln\frac{\bar\Lambda}{\mu}\qquad\nonumber\\
&& +\frac{1}{4}\,\mbox{tr}\, C(R)\, \Big(3C_2- 2 N_f T(R)\Big)\Big(\ln^2\frac{\bar\Lambda}{\mu}+\frac{2}{\varepsilon}\ln\frac{\bar\Lambda}{\mu} + \frac{3}{2}\ln\frac{\bar\Lambda}{\mu}\Big) \Big] + \varepsilon\mbox{-poles} + O(\alpha_{s0}^3)\Big\}.
\end{eqnarray}

\noindent
Here we take into account that this expression also contains some $\varepsilon$-poles (with the coefficients depending on $\alpha_{s0}$) which vanish after differentiating with respect to $\ln\mu$. To find these $\varepsilon$-poles we note that the relation between $\alpha^{-1}$ and $\alpha_0^{-1}$ is constructed by requiring finiteness of the function $d^{-1}$ expressed in terms of the renormalized quantities. In the case of using the regularization by dimensional reduction this function can be written in the form

\begin{equation}\label{Dinverse}
d^{-1}_{\overline{\mbox{\scriptsize DR}}}=\alpha_0^{-1} + \big(\Lambda_{\mbox{\scriptsize DR}}\big)^{\varepsilon} I_1 + \alpha_{s0} \big(\Lambda_{\mbox{\scriptsize DR}}\big)^{2\varepsilon} I_2 + \alpha_{s0}^2 \big(\Lambda_{\mbox{\scriptsize DR}}\big)^{3\varepsilon} I_3 + O(\alpha_{s0}^3) + O(\alpha_0),
\end{equation}

\noindent
where $I_1$, $I_2$, and $I_3$ are the one-, two-, and three-loop integrals, respectively. They have the structure

\begin{eqnarray}\label{I_Expansion}
&&\big(\Lambda_{\mbox{\scriptsize DR}}\big)^{\varepsilon}I_1 = \frac{1}{\pi}\sum_{\alpha}q_{\alpha}^2 \Big(\frac{\bar\Lambda}{P}\Big)^{\varepsilon} \Big(\frac{c_{1,1}}{\varepsilon}+c_{1,0}\Big);\nonumber\\
&&\big(\Lambda_{\mbox{\scriptsize DR}}\big)^{2\varepsilon} I_2 =
\frac{1}{\pi^2}\sum_{\alpha}q_{\alpha}^2 \Big(\frac{\bar\Lambda}{P}\Big)^{2\varepsilon} \Big(\frac{c_{2,1}}{\varepsilon}+c_{2,0}\Big);\nonumber\\
&&\big(\Lambda_{\mbox{\scriptsize DR}}\big)^{3\varepsilon}I_3 =
\frac{1}{\pi^3}\sum_{\alpha}q_{\alpha}^2 \Big(\frac{\bar\Lambda}{P}\Big)^{3\varepsilon} \Big(\frac{c_{3,2}}{\varepsilon^{2}}+\frac{c_{3,1}}{\varepsilon}+c_{3,0}\Big),
\end{eqnarray}

\noindent
where $c_{i,j}$ are some numerical coefficients. (Expanding $(\bar\Lambda/P)^{n\varepsilon}$ we obtain terms containing $\ln\bar\Lambda/P$). Substituting $\alpha_0^{-1}$ from Eq. (\ref{Alpha_Relation_Original}), $\alpha_{s0}$ from Eq. (\ref{AlphaS_Relation_DR}) and requiring that all $\ln\bar\Lambda$ vanish, we find the coefficients

\begin{eqnarray}\label{C_Coefficients}
&& c_{1,1} = \dim(R);\qquad c_{2,1}= \frac{1}{2}\,\mbox{tr}\, C(R);\qquad c_{3,2} = \frac{1}{6} \Big(3C_2 - 2 N_f T(R)\Big)\, \mbox{tr}\, C(R);\qquad\nonumber\\
&&\qquad\quad c_{3,1} = \Big(3 C_2 - 2 N_f T(R)\Big) \Big(\frac{1}{2}c_{2,0} + \frac{1}{8}\mbox{tr}\, C(R)\Big) - \frac{1}{6} \mbox{tr}\left(C(R)^2\right).
\end{eqnarray}

\noindent
Note that the coefficient $c_{3,1}$ cannot be completely found in this way, because (due to the presence of $1/\varepsilon$ in Eq. (\ref{AlphaS_Relation_DR})) it appears to be related to $c_{2,0}$, which should be calculated separately. Substituting the values of $c_{i,j}$ from Eq. (\ref{C_Coefficients}) into Eqs. (\ref{Dinverse}) and (\ref{I_Expansion}) we obtain $\varepsilon$-poles in the function $d^{-1}_{\overline{\mbox{\scriptsize DR}}}$ which should be cancelled by the $\varepsilon$-poles in the function (\ref{Alpha_Relation_Original}). This gives the relation between the bare and SQCD-renormalized electromagnetic coupling constant,

\begin{eqnarray}\label{Alpha_Relation}
&&\hspace*{-7mm} \frac{1}{\alpha_0} = \frac{1}{\alpha} - \frac{1}{\pi} \sum_{\alpha} q_{\alpha}^2 \Big\{\dim(R)\Big(\frac{1}{\varepsilon}+\ln\frac{\bar\Lambda}{\mu}\Big)+\frac{\alpha_{s}}{\pi} \mbox{tr}\, C(R)\Big(\frac{1}{2\varepsilon}+\ln\frac{\bar\Lambda}{\mu}\Big) + \frac{\alpha_s^2}{\pi^2}\, \Big[ -\frac{1}{2}\mbox{tr}\left(C(R)^2\right) \nonumber\\
&&\hspace*{-7mm} \times \Big(\frac{1}{3\varepsilon}+\ln\frac{\bar\Lambda}{\mu}\Big) -\frac{1}{4} \mbox{tr}\, C(R) \Big(3C_2 - 2 N_f T(R)\Big)\Big(\frac{1}{3\varepsilon^2}+\frac{1}{\varepsilon}\ln\frac{\bar\Lambda}{\mu} +\ln^2\frac{\bar\Lambda}{\mu} - \frac{1}{2\varepsilon} - \frac{3}{2}\ln\frac{\bar\Lambda}{\mu}   \Big) \Big]\nonumber\\
&&\hspace*{-7mm} + O(\alpha_{s}^3)\Big\}.\vphantom{\frac{1}{2}}
\end{eqnarray}

\noindent
Next, we substitute $\alpha_0^{-1}$ from (\ref{Alpha_Relation}) into Eq. (\ref{D_Function_Bare}) and rewrite the result in terms of the bare SQCD coupling constant $\alpha_{s0}$. This gives the three-loop $D$-function defined in terms of the bare coupling constant for the theory regularized by the dimensional reduction,

\begin{eqnarray}
&& D(\alpha_{s0})=\frac{3}{2}\sum_{\alpha}q_{\alpha}^2\Big\{\mbox{dim}(R) + \frac{\alpha_{s0}}{\pi}\, \mbox{tr}\,C(R) + \frac{\alpha_{s0}^2}{\pi^2}\, \Big[-\frac{1}{2}\mbox{tr} \left(C(R)^2\right)\nonumber\\
&&\qquad\qquad\qquad\qquad\qquad\qquad\quad +\Big(\frac{3}{2}+\frac{1}{\varepsilon}\Big) \mbox{tr}\,C(R)\, \Big(\frac{3}{4}C_2-\frac{1}{2}N_fT(R)\Big)\Big] + O(\alpha_{s0}^3)\Big\}.\qquad
\end{eqnarray}

Also we need the anomalous dimension defined in terms of the bare coupling constant for the theory regularized by the dimensional reduction. It is calculated by the same method. We start with the expression (\ref{Anomalous_Dimension}) (in which $\alpha_s$ should be expressed in terms of $\alpha_{s0}$ with the help of Eq. (\ref{AlphaS_Relation_DR})) and substitute it into the renormalization group equation

\begin{equation}
\widetilde\gamma(\alpha_s)_i{}^j = \left.\frac{d}{d\ln\mu}\ln Z(\alpha_s,\ln\bar\Lambda/\mu,1/\varepsilon)_i{}^j\right|_{\alpha_{s0}=\mbox{\scriptsize const}}.
\end{equation}

\noindent
Integrating with respect to $\ln\mu$ and rewriting the result in terms of the renormalized SQCD coupling constant $\alpha_s$ with the help of Eq. (\ref{AlphaS_Relation_DR}) we obtain

\begin{eqnarray}
&& \ln Z_i{}^j = \frac{\alpha_{s0}}{\pi} C(R)_i{}^j \ln\frac{\bar\Lambda}{\mu} + \frac{\alpha_{s0}^2}{\pi^2}\Big[\, \frac{1}{4}\Big(3 C_2 - 2 N_f T(R)\Big) C(R)_i{}^j \Big(\ln^2\frac{\bar\Lambda}{\mu} +\frac{2}{\varepsilon} \ln\frac{\bar\Lambda}{\mu} - \ln\frac{\bar\Lambda}{\mu} \Big)\qquad\nonumber\\
&& -\frac{1}{2}\left(C(R)^2\right)_i{}^j \ln\frac{\bar\Lambda}{\mu} \Big] +\varepsilon\mbox{-poles} + O(\alpha_{s}^3),
\end{eqnarray}

\noindent
where the coefficients of the $\varepsilon$-poles depend on $\alpha_{s0}$. To find these $\varepsilon$-poles, we note that the expression $\ln Z_i{}^j + \ln G_i{}^j$ should be finite, while the function $\ln G_i{}^j$ is given by the sum of Feynman diagrams of the structure

\begin{equation}
\ln G_i{}^j =  \alpha_{s0} \big(\Lambda_{\mbox{\scriptsize DR}}\big)^{\varepsilon} \big(G_1\big)_i{}^j + \alpha_{s0}^2 \big(\Lambda_{\mbox{\scriptsize DR}}\big)^{2\varepsilon} \big(G_2\big)_i{}^j + O(\alpha_{s0}^3),
\end{equation}

\noindent
where

\begin{eqnarray}\label{G_Expansion}
&&\big(\Lambda_{\mbox{\scriptsize DR}}\big)^{\varepsilon} \big(G_1\big)_i{}^j = \frac{1}{\pi} \Big(\frac{\bar\Lambda}{P}\Big)^{\varepsilon} \Big(\frac{1}{\varepsilon} \big(k_{1,1}\big)_i{}^j + \big(k_{1,0}\big)_i{}^j \Big);\nonumber\\
&&\big(\Lambda_{\mbox{\scriptsize DR}}\big)^{2\varepsilon}\big(G_2\big)_i{}^j =
\frac{1}{\pi^2} \Big(\frac{\bar\Lambda}{P}\Big)^{2\varepsilon} \Big(\frac{1}{\varepsilon^2}\big(k_{2,2}\big)_i{}^j+\frac{1}{\varepsilon}\big(k_{2,1}\big)_i{}^j + \big(k_{2,0}\big)_i{}^j\Big).
\end{eqnarray}

\noindent
From Eqs. (\ref{Z_DR}) and (\ref{G_Renormalized_DR}) we conclude that

\begin{equation}
\big(k_{1,1}\big)_i{}^j = \big(k_{1,0}\big)_i{}^j = - C(R)_i{}^j.
\end{equation}

\noindent
The same value of $\big(k_{1,1}\big)_i{}^j$ is certainly obtained from the requirement that the one-loop contribution to the function $\big(G_{R,\overline{\mbox{\scriptsize DR}}}\big)_i{}^j$ does not contain $\ln\bar\Lambda$. The similar requirement for the two-loop contribution gives

\begin{eqnarray}
&& \big(k_{2,2}\big)_i{}^j = -\frac{1}{4} \Big(3C_2 - 2 N_f T(R)\Big) C(R)_i{}^j;\nonumber\\
&& \big(k_{2,1}\big)_i{}^j = \frac{1}{4} \big(C(R)^2\big)_i{}^j - \frac{5}{8} \Big(3 C_2 - 2 N_f T(R)\Big) C(R)_i{}^j.\quad
\end{eqnarray}

\noindent
When these values of the coefficients $\big(k_{m,n}\big)_i{}^j$ are known, then it is possible to find all $\varepsilon$-poles in the function $\ln G_i{}^j$. They should be canceled by the $\varepsilon$-poles in the function $\ln Z_i{}^j$. Therefore, we are able to find the complete expression for this renormalization constant,

\begin{eqnarray}
&&\hspace*{-5mm} \ln Z_i{}^j = \frac{\alpha_{s}}{\pi} C(R)_i{}^j \Big(\frac{1}{\varepsilon} + \ln\frac{\bar\Lambda}{\mu}\Big) + \frac{\alpha_{s}^2}{\pi^2}\Big[-\frac{1}{4}\Big(3 C_2 - 2 N_f T(R)\Big) C(R)_i{}^j \Big(\,\frac{1}{\varepsilon^2} +\frac{2}{\varepsilon} \ln\frac{\bar\Lambda}{\mu}  +\ln^2\frac{\bar\Lambda}{\mu} \nonumber\\
&&\hspace*{-5mm} - \frac{1}{2\varepsilon} - \ln\frac{\bar\Lambda}{\mu} \Big)
-\frac{1}{2}\left(C(R)^2\right)_i{}^j \Big(\frac{1}{2\varepsilon} + \ln\frac{\bar\Lambda}{\mu}\Big) \Big] + O(\alpha_{s}^3).
\end{eqnarray}

\noindent
Differentiating the result with respect to $\ln\bar\Lambda$ gives the anomalous dimension defined in terms of the bare coupling constant,

\begin{eqnarray}
&& \gamma(\alpha_{s0})_i{}^j = \left. - \frac{d}{d\ln\bar\Lambda}\ln Z_i{}^j\right|_{\alpha_s=\mbox{\scriptsize const}}\nonumber\\
&&\qquad = -\frac{\alpha_{s0}}{\pi} C(R)_i{}^j + \frac{\alpha_{s0}^2}{\pi^2}\Big[-\frac{1}{4}\Big(3 C_2 - 2 N_f T(R)\Big) C(R)_i{}^j  +\frac{1}{2} \big(C(R)^2\big)_i{}^j\Big] + O(\alpha_{s0}^3),\qquad
\end{eqnarray}

\noindent
where we took into account that the result should be expressed in terms of the bare SQCD coupling constant $\alpha_{s0}$.

\end{document}